\documentclass{article}
\usepackage[utf8]{inputenc}

\usepackage{natbib}
\usepackage{graphicx}
\usepackage{graphicx,color}
\usepackage{amsmath}
\usepackage{amssymb}
\usepackage{graphicx}
\usepackage{epstopdf}
\usepackage{inputenc}
\usepackage{geometry}
\usepackage{float}
\usepackage{ulem}
\usepackage{longtable}
\usepackage[colorlinks=true,linkcolor=blue,citecolor=blue]{hyperref}
\usepackage{fancyhdr}
\usepackage{pgfplots}
\usepackage{booktabs}
\usepackage{natbib}
\usepackage{color}
\usepackage{subfigure}
\usepackage{multirow}
\usepackage{threeparttable}
\usepackage[linesnumbered,ruled,vlined]{algorithm2e}
\DontPrintSemicolon

\title{Lapse risk modelling in insurance:
a Bayesian mixture approach}

\author{Viviana G. R. Lobo$^{1}$\footnote{{{\it Corresponding author}: Viviana G R Lobo, Departamento de M\'etodos Estat\'{\i}sticos, Instituto de Matem\'atica, Universidade Fe\-de\-ral do Rio de Janeiro, Av. Athos da Silveira Ramos, Centro de Tecnologia, Bloco C, CEP 21941-909. \newline {\it E-mail}: {\tt viviana@dme.ufrj.br}. {\it Homepage}: https://sites.google.com/site/dme/viviana}} ,  Thais C. O. Fonseca$^{1}$, Mariane B. Alves$^{1}$  \\
\textit{$^{1}$Instituto de Matem\'atica,} \\ 
\textit{Universidade Federal do Rio de Janeiro, Brazil }
}

\date{}

\begin{document}

\maketitle

\begin{abstract}
    This paper focuses on modelling surrender time for policyholders in the context of life insurance. In this setup, a large lapse rate at the first months of a contract is often observed, with a decrease in this rate after some months. The modelling of the time to cancellation must account for this specific behaviour. Another stylised fact is that policies which are not cancelled in the study period are considered censored. To account for both censuring and heterogeneous lapse rates, this work assumes a Bayesian survival model with a mixture of regressions. The inference is based on data augmentation allowing for fast computations even for data sets of over a million clients. Moreover, scalable point estimation based on EM algorithm is also presented. An illustrative example emulates a typical behaviour for life insurance contracts and a simulated study investigates the properties of the proposed model. In particular, the observed censuring in the insurance context might be up to $50\%$ of the data, which is very unusual for survival models in other fields such as epidemiology. This aspect is exploited in our simulated study. 
\end{abstract}


\section{Introduction}

\subsection{Background}
{Lapse rate risk modelling is an important issue that  is getting attention from life insurance markets. In the context of life insurance, this is an even more important issue, since  contracts have longer policy term and large rates of surrender. Originally, the term lapse means termination of an insurance policy and loss of coverage because the policyholder has failed to pay premiums (\cite{Gazert2009}; \cite{Kuo2003}, \cite{Eling2013b}). In this paper, lapse risk refers to the life policies surrendered before their maturity or cancelled contracts when the policyholder fails to comply with their obligations (e.g. premium payment). In  other words, when a customer cancels their policy or surrenders this policy, either to switch insurance companies or because someone is no longer interested, we consider that the customer has churned. 

Due to the large impact lapses may produce on an insurer’s portfolio, particularly in the first periods of the contracts, it is important to understand the factors that drive its risk. Large changes in lapse rates 
can potentially lead to financial losses which can prevent insurers from complying with their contractual obligations. Furthermore, lapse rates can  be difficult to model due to the fact that while doing so, one needs not only  to take into account the policyholder’s behavioural features, but also the characteristics of the life insurance products being acquired. Once factors associated with cancellation or surrender are identified, customer retention programmes can be developed and actions can be taken \citep{Gunther2014}. Moreover, good  persistence  is  of  vital  importance  to  the  financial  performance  of  life  insurance companies.

We aim to address some issues related to churn, such as the existence of trends in the persistence of specific products or groups of products. Those factors may enhance the persistence curve in the insurance company. In addition, churn/lapsing impacts many actuarial tasks, such as product design, pricing, hedging and risk management.}

 The rate of cancellation varies according to the product and profile of policyholders. A statistical model can be used to identify the risk factors affecting persistency (or lapse) rate over time or for pricing new business taking the cancellation risk into account. For instance, the model could identify the products with highest risk of cancellation. The main statistical models usually considered in this area are: hierarchical regression models, survival regression models and time series models. \cite{Mil18} considers a competing risk approach in the context of a survival regression models. \cite{Eling2013} consider the proportional hazard models and generalized linear models to show that product characteristics such as product type or contract age and policyholder characteristics are important drivers for lapse rates illustrated by a data set provided by a German life insurer. \cite{brockett2008} pay particular attention to household customer behaviour considering households for whom at least one policy has lapsed and investigates the effects of the lapse rates of other policies owned by the same household. They use the logistic regression and survival analysis techniques to assess the probability of total customer withdrawal, and the length of time between first cancellation and subsequent customer withdrawal. In the literature, other authors consider the same techniques presented in \cite{brockett2008} to analyse customer churn. 
 \cite{Gunther2014} present a dynamic modelling approach for predicting individual customers’ risk of lapse. They consider a logistic longitudinal regression model that incorporates time-dynamic explanatory variables and interactions is fitted to the data. Our work is based on the experience gained from observing higher cancellation rates at the beginning of the contracts, decreasing after some months.  Issues such as if there is a trend of persistency in a specific product or group of products, and if there are factors which can enhance the persistency curve in the company, will be investigated. Possibly, these factors can be controlled by the insurer leading to increased persistency.


Specifically, interest lies, in this work, on modelling the time to cancellation of a contract or surrender. This includes the contracts which are terminated by the policyholder or terminated by the insurer due to lack of premium payment and does not include external events such as death. Policies which are not considered cancelled are defined as censured, as the actual time to cancellation has not yet been observed in the study period.


 We follow the Bayesian paradigm to model the survival time of a policy via parametric mixtures of survival regression models extending usual survival approaches to censured data (\cite{Ibrahim2001}, \cite{Kalbfleisch2002}), flexibilizing survival and failure rate curves. Inference is performed via Gibbs sampler and other Markov chain Monte Carlo (MCMC) techniques, fitting particular survival models especially in the presence of complex censoring schemes. 
We propose a flexible model able to deal with data sets of thousands of policyholders. In this context, the inference is based on data augmentation allowing for fast and feasible computations (see \cite{tanner1987calculation}) and scalability is achieved by the adoption of Expectation Maximization (EM) algorithms. With regard to flexibilization of survival curves so that they can capture higher cancellation rates at the beginning of the term, \cite{McLachlan1994} consider finite mixture models to analyse failure-time data in a variety of situations. Our proposed model allows to accommodate heterogeneous behaviours in the lapse rates via a Bayesian mixture models based on \cite{fruhwirth2006finite} taking into account a large volume of censored data. 

\subsection{An illustration via usual parametric survival models}\label{sec1.2}
To illustrate the larger surrender rates at the first months of a contract and a smaller rate later in time we consider an example and assume an usual survival model, which proves to be inadequate for this kind of insurer portfolio behaviour. Let $T_i$ be the non-negative random variable denoting the duration of a policy $i$ before termination (cancellation). Note that $T_i$ could be modelled using a survival function given by
\begin{equation}
    S(t)= P(T_i > t) = \int_{t}^{\infty} f(u) du,
\end{equation}

\noindent where $S(t)$ is monotonic and decreasing function, starting at $S(0) = 1$ and converging numerically to zero, since $S(\infty) = \lim\limits_{t \to \infty} S(t) = 0$, and $f(t)$ denotes the probability density function of $T_i$. The cancellation rate, $\lambda(t)$, is given by

\begin{equation}\label{eq:taxafalha1}
\lambda(t) = \lim_{\Delta t \to 0} \frac{P(t<T_i\leq t+ \Delta t\mid T_i>t)}{\Delta t} \approx \frac{f(t)}{S(t)} = -\frac{S'(t)}{S(t)},
\end{equation}

\noindent where $\Delta t$ is a small-time increment. In particular, $\lambda(t)\Delta t$ is the approximate probability of a failure occurring in the interval $(t,t+\Delta t)$, i.e., the lapse rate, given that it has survived until time $t$ \citep[more details see][]{Ibrahim2001}.

In many situations, data collected in the context of failure  times contain observations that are censored. 
In our context, we define censored data as  current policies, occurrences of  claims (death of the insured), and terminations of contracts.

An artificial database was simulated in order to emulate the cancellation behaviour in insurance products with 1,000 policies, 40\% censored data and considering a dichotomous covariate $x$ (0/no attribute, 1/yes attribute). We let the observed failure-time data be denoted by
 \begin{equation}
   d_i= (t_i, \delta_i, x_i), \quad  i=1, \ldots, 1,000.
 \end{equation}
 \noindent where the $t_i$ is the time recorded for the {\it i}th policy, $x_i$ is a covariate associated with the {\it i}-th policy, and $\delta_i$ is an indicator of the censoring status, given by
      \[ \delta_i =
  \begin{cases}
   1       &  \text{if} \quad t_i \quad \text{is a failure time}\\
   0  & \text{if} \quad t_i  \quad \text{is a censored time}.
  \end{cases}
\] 

Thus, $\delta_i=1$ is the event representing the lapse risk (with $f(t_i)$ distribution), whereas $\delta_i= 0$ (with $S(t_i)$ survival), the lapse risk time is known only to be greater than $t_i$, i.e., the survival time is censored. The observations are assumed to be independent for different policies.

\begin{figure}[H]
    \centering 
    \begin{tabular}{cc}
          \includegraphics[width=5cm]{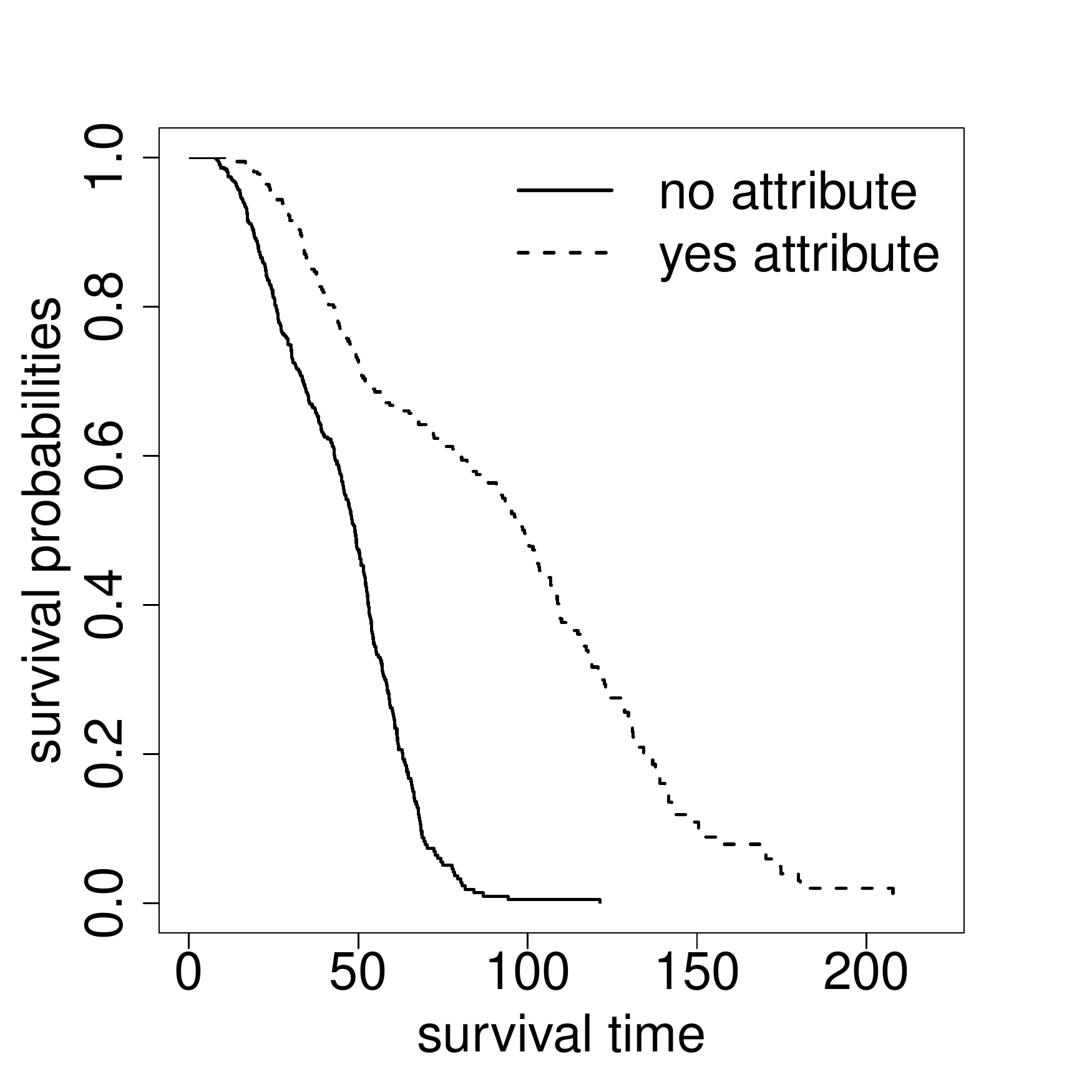} &   \includegraphics[width=5cm]{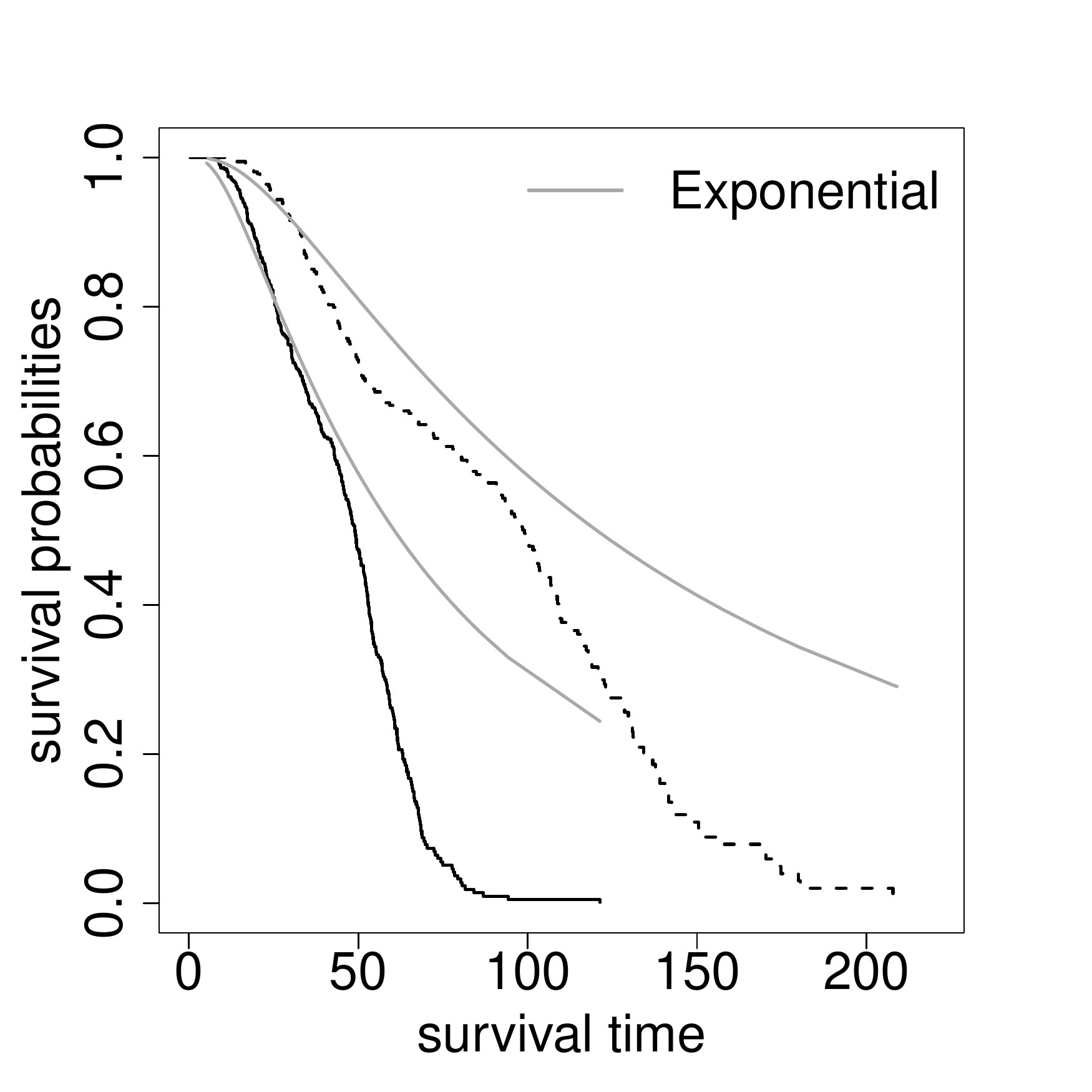}\\
    (a) Empirical survival curve  & (b) Exponential fit \\ 
       \includegraphics[width=5cm]{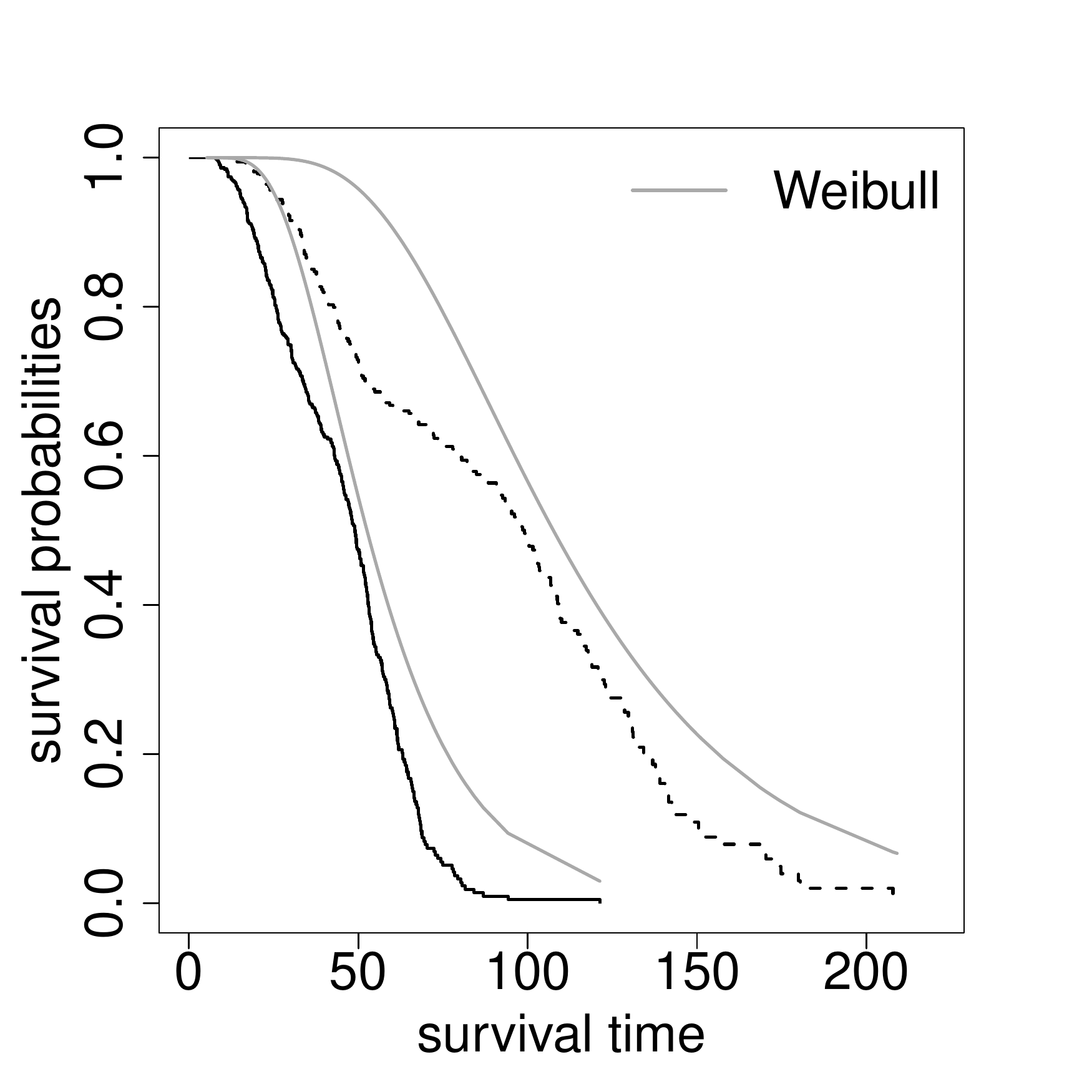} &   \includegraphics[width=5cm]{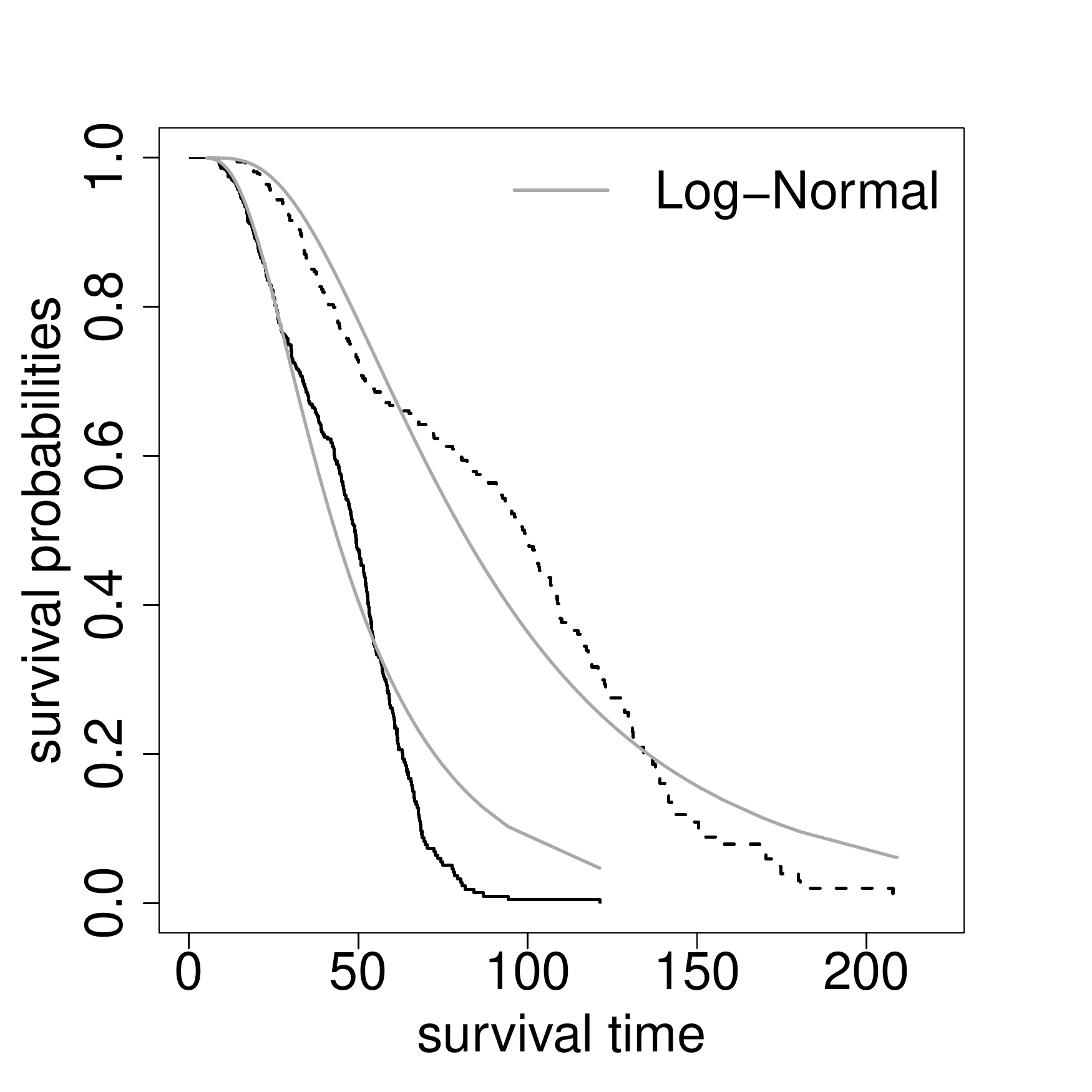}\\
    (c) Weibull fit & (d) Log-normal fit \\ 
    \end{tabular}
    \caption{Simulated data set: (a) Eempirical Kaplan-Meier survival curves, (b) Exponential, (c) Weibull and (d) Log-Normal fitted models.}
    \label{fig1}
\end{figure}

Figure \ref{fig1} (a) presents the empirical Kaplan-Meier survival curve for both attributes and it is clear that there is an heterogeneous behaviour between the levels of the dichotomous covariate. The absence of the attribute described by the covariate is associated with increased premature risk of cancellation.  In addition, for both levels of the covariate, there is a difference in the survival behaviour in the initial times when compared to the following ones. 

In the context of survival analysis, parametric models play a key role in modelling the phenomenon of interest. If  $T_i$ follows an Exponential model, then $\lambda(t) = \lambda$ is constant over
time. If a Weibull model with parameters $\lambda$ and $\kappa$ is considered for $T_i$, then $\lambda(t)= \lambda \kappa t^{\kappa -1}$. The Log-Normal model with parameters $\mu$ (mean logarithm of the failure time) and $\sigma$ (standard deviation) is suitable when the interest is that the failure rate is not monotonous, but it reaches a maximum point and then decreases. A regression model could be considered with covariates relating the parameters in the sampling model with covariates for each contract. 

Panels (b)--(d) in Figure \ref{fig1} make it clear that usual survival models are not flexible to accommodate different phases of $T_i$ over time. Even though the Log-Normal model produces a good performance when compared with the competing models in the initial instants, it fails to adapt in later times. That is, the rates tend to slow down over time but usual parametric survival models are not able to accommodate this behaviour. Our working premise is that simple parametric models can serve as block builders of more flexible models, via mixtures.

\subsection{Outline of the paper}
The remaining of the paper is organised as follows. Section \ref{sec2} describes the proposed model and its properties. In particular, Sections \ref{sec2.2} and \ref{sec2.3} describe the inference and computational procedures for mixture survival modelling via data augmentation and Section \ref{EM} describes the adoption of an EM algorithm that makes the inferential process scalable. Section \ref{sec3} presents the application of the proposed methods considering two simulated data sets. The first one studies effectiveness of our proposal in modelling survival curves that have different behaviours over time with low computational cost and the following one models mixture survival curves for an insurance company through an artificial data set, aiming to obtain feasible results via our proposed model. Furthermore,  the formulation enables to compute the churn probability, in specified time intervals, for groups of policyholders sharing similar features. Section \ref{sec4} concludes with final discussion and remarks. Some aspects about the simulated data sets are presented in Appendix \ref{apB}.


\section{Bayesian mixture survival model}\label{sec2}

In this section, we propose a mixture of parametric models for censored survival data. Finite mixture models are described in detail in  \cite{fruhwirth2006finite}. As seen in the illustration with artificial data presented in Section \ref{sec1.2}, the competing fitted models apparently do not reflect the empirical distribution of the data. A possible alternative is to use more flexible structures such as mixture models, which allow the incorporation of behavioural change in the probability distribution of the data. Our proposal is based on the classical finite mixture model, where observations are assumed to arise from the mixture distribution given by
\begin{equation}
f(t_i) =     \sum_{j=1}^{K} \eta_j f_j(t_i),
\end{equation}
\noindent where $f(t_i)$ is the probability density function of $T_i$ and $f_j(t_i)$ denotes  $K$ component densities occurring with unknown proportions $\eta_j$, with $0 \leq \eta_j \leq 1$ and $\sum_{j=1}^{K} \eta_j=1$. It follows that the survival function $S(t)$ has the  mixture form
\begin{equation}
    S(t) = \sum_{j=1}^{K} \eta_j S_j(t) =   \sum_{j=1}^{K} \eta_j \int_{t}^{\infty} f_j(u)du, \quad j=1, \ldots, K.
\end{equation}
For each policy $i$, we define a latent group indicator $I_i  \mid \boldsymbol{\eta} \sim Categorical(K, \boldsymbol{\eta})$, with $\boldsymbol{\eta}=(\eta_1, \ldots, \eta_K)$, following a categorical distribution given by $P(I_i \mid \boldsymbol{\eta})= \prod_{j=1}^{K} [\eta_j]^{I_{ij}}$,  where $I_{ij}=1$  if  observation $i$ is allocated to group $j$ and is null, otherwise. These auxiliary non-observable variables aim to identify which mixture component each observation has been generated from and their introduction in the formulation makes it simple to express the likelihood function for observation $i$:
\begin{equation}\label{eq6}
    f_{T_i}(t_i \mid I_i) = \prod_{j=1}^{K} [f_j(t_i)]^{I_{ij}}, \quad i=1, \ldots, n.
\end{equation}

In this work, the number $K$ of mixture components is assumed to be  known. Notice that in the example shown in Section \ref{sec1.2}, $K=2$ components are assumed to model the lapse risk. In this case, the lapse risk can be decomposed into two overlapping processes in time. The first process is associated to the period that immediately follows the contracting of the policy. In general it covers the first three months, when the lapse risk  is relatively high. The second process refers to the subsequent period, when persistence decreases smoothly over time. Thus, the mixture  distribution is written as $f_{T_i}(t_i \mid I_i) = [f_1(t_i)]^{I_{i1}}[f_2(t_i)]^{(1-I_{i1})}$.

\subsection{Log-Normal mixture survival model}\label{sec2.1}
In this paper, interest lies in modelling $Y_i = log(T_i)$, the logarithmic duration of a policy $i$ before termination (cancellation), such that $Y_i \sim \mathcal{N}(\mu_i,\sigma^2)$, implying that in the original scale  $T_i \sim \mathcal{LN}(\mu_i,\sigma^2)$, $i=1, \ldots, n$. Survival times associated to a different outcome than the lapse, such as survival times past the end of our study and deaths, are assumed to be censored for policies, $i= h+1, \ldots, n$. , then 
\begin{equation}\label{eq:lognormal}
f(t_i\mid \mu, \sigma^2) = (2\pi)^{-\frac{1}{2}}(t_i\sigma)^{-1}\exp\left\{-\frac{1}{2\sigma^2}(\log(t_i)-\mu_i,)^2\right\}.
\end{equation}
The resulting survival function is given by
\begin{equation}\label{eq:sobrevivencialognormal}
S(t_i \mid \mu_i, \sigma^2)= 1 - \Phi\left(\frac{\log(t_i)- \mu_i}{\sigma}\right).
\end{equation}
We can thus write the survival likelihood function  of $(\mu, \sigma^2)$ implied by a log-normal model and based on data $D$ as 
\begin{equation}\label{eq9}
 L(\mu_i, \sigma^2 \mid D) = \prod_{i=1}^{n} f(t_i \mid \mu_i, \sigma^2)^{\delta_i} S(t_i \mid \mu_i, \sigma^2)^{(1- \delta_i)}.  
\end{equation}
Adopting the finite mixture approach and assuming  $Y_i= log(T_i) \mid \mu_i, \sigma^2 \sim \mathcal{N}(\mu_i, \sigma^2)$ and latent variables $I_{ij}$, $i=1,\ldots, n$,  it follows that:
\begin{equation}\label{eq:mixture2}
    f_{Y_i}\left(y_{i} \mid \mathrm{I}_{i}, \mu_i, \sigma^2 \right)=
    \prod_{j=1}^{K} {[\mathcal{N}_j}(y_i \mid \mu_{ij}, \sigma^2_j)]^{\mathrm{I}_{ij}},
\end{equation}
\noindent where $\mathcal{N}_j$ is a  normal distribution for the component group $j$, $j=1,\ldots,K$. Then,
\begin{eqnarray}\label{eq:mixture3} \nonumber
{y}_{i}\mid \{\mathrm{I}_{i1}=1 \}, \boldsymbol{\beta}_1, \sigma^2_1 &\sim& \mathcal{N}_{1}(\mu_{i1}(\boldsymbol{\beta}_1),\sigma^2_{1}) \\ \nonumber
{y}_{i}\mid \{\mathrm{I}_{i2} = 1 \}, \boldsymbol{\beta}_2, \sigma^2_2 &\sim& \mathcal{N}_{2}(\mu_{i2}(\boldsymbol{\beta}_2),\sigma^2_{2}) \\ \nonumber
\vdots && \vdots \\ 
{y}_{i}\mid \{\mathrm{I}_{iK}=1 \}, \boldsymbol{\beta}_K, \sigma^2_K &\sim& \mathcal{N}_{K}(\mu_{iK}(\boldsymbol{\beta}_K),\sigma^2_{K}),  \nonumber
\end{eqnarray}
\noindent with $\mu_{ij}= \mathbf{x}_{ij}^T \boldsymbol{\beta}_j$,  $\boldsymbol{\beta}_j = (\beta_{0j}, \beta_{1j}, \ldots, \beta_{pj})$ characterizing the unknown mean and $ \sigma^2_{j}$, the variance, respectively, for $j=1,\ldots, K$ and $i=1,\ldots, n$. 

Using the latent indicators of categorical allocation, the likelihood simplifies to
\begin{eqnarray}\nonumber
f(\mathbf{y}\mid \boldsymbol{\eta}, \boldsymbol{\beta}, \sigma^2) &=&  \prod_{j=1}^{K}\prod_{i=1}^{n}\eta_j^{I_{ij}} [\mathcal{N}_j(y_i \mid \boldsymbol{\beta}_{j}, \sigma^2_{j})]^{I_{ij}} \\ 
 &=& \prod_{j=1}^{K} \eta_j^{n_j} \left[ \prod_{i: I_{ij}=1} \mathcal{N}_j(y_i \mid \boldsymbol{\beta}_{j}, \sigma^2_j)\right], 
\end{eqnarray}
\noindent where $n_j=\sum_{i}I_{ij}$ is the number of observations allocated to group $j$ and $n= \sum_{j=1}^{K}n_j$, for $j=1,\ldots, K$,  $i=1,\ldots, n$. Thus, the mixture survival likelihood function is given by
\begin{equation}\label{eq12}
    f(\mathbf{y} \mid \boldsymbol{\eta}, \boldsymbol{\beta}, \sigma^2) = \prod_{j=1}^{K} \eta_j^{n_j} \left[ \prod_{i: I_{ij}=1} \mathcal{N}_j(y_i \mid \boldsymbol{\beta}_{j}, \sigma^2_j)^{\delta_i} S(y_i \mid \boldsymbol{\beta}_j, \sigma^2_j)^{1-\delta_i}\right] , 
\end{equation}
\noindent where $\delta_i$ in the censorship indicator, as previously seen in Section \ref{sec1.2}. 

 From a Bayesian point of view, we are interested in the posterior $p(\boldsymbol{\eta}, \boldsymbol{\beta}, \sigma^2 \mid \mathbf{y})$. The posterior distribution are generally not available analytically, and numerical integration and simulation are considered, in particular, Markov chain Monte Carlo (MCMC) methods \citep{gamerman} are used in this paper. Notice that to compute posterior distributions, we need to take into account the censored quantities, which in practice can be computationally prohibitive, depending on the percentage of censored data and the large data set. Thus, inference is facilitated through data augmentation.

\subsection{Inference based on data augmentation}\label{sec2.2}
The  presence of censored data is a common feature when considering time data until the occurrence of an event and the likelihood function takes this fact into account,  as seen in equation (\ref{eq9}). Following the Bayesian approach, the estimation procedure can be based on MCMC algorithm using the data augmentation technique \citep[see][]{tanner1987calculation}. 
Suppose we observe survival times $\mathbf{y}^{obs}= (y^{obs}_1, \ldots, y^{obs}_h)$. Then the idea is to define the survival times for the $n-h$ censored policies as missing data which we denote as $\mathbf{z}= (z_{h+1}, \ldots, z_{n})$.

Assume that the full data is given by
 \begin{equation}
     \mathbf{y}= (\mathbf{y}^{obs}, \mathbf{z}).
 \end{equation}
Let $\boldsymbol{\theta}$ the parametric vector of interest. The data augmentation approach is motivated by the following representation of the posterior density
 \begin{equation}
     p(\boldsymbol{\theta} \mid \mathbf{y}^{obs}, \boldsymbol{\delta}) = \int_{\mathbf{Z}} p(\boldsymbol{\theta} \mid \mathbf{y}^{obs}, \mathbf{z}, \boldsymbol{\delta})p(\mathbf{z} \mid \mathbf{y}^{obs}, \boldsymbol{\delta}) d\mathbf{z},
 \end{equation}
 \noindent where the vector $\boldsymbol{\delta}$ is composed by censorship indicators $\delta_i  \in \left\{0, 1\right\}$; $p(\boldsymbol{\theta} \mid \mathbf{y}^{obs},\boldsymbol{\delta})$ denotes the posterior density of the parameter $\boldsymbol{\theta}$ given the observed data $\mathbf{y}^{obs}$; $p(\mathbf{z} \mid \mathbf{y}^{obs}, \boldsymbol{\delta})$ denotes the predictive density of latent data $\mathbf{z} $ given $\mathbf{y}^{obs}$; and $p(\boldsymbol{\theta} \mid \mathbf{y}^{obs}, \mathbf{z} , \boldsymbol{\delta})$ the conditional density of $\boldsymbol{\theta}$ given the augmented data $\mathbf{y}$.
 
 In practice, we do not know {\it a priori} to  which group a given observation belongs. Thus, in addition to the censored observations, whose outcome is unknown, the variable $I_i$ in equation (\ref{eq6}) is also latent and is estimated in our inferential algorithm. Assuming that policies are independent, the likelihood function for the complete data can be written as
 \begin{eqnarray}\label{eq15}
 f(\mathbf{y}^{obs}, \mathbf{z}, \boldsymbol{\delta} \mid \boldsymbol{\theta}) 
     &=& { \prod_{j=1}^{K} \eta_j^{n_j}} \left[\prod_{i : \delta_i=1, I_{ij}=1}f{_{j}}(y_i^{obs} \mid \boldsymbol{\theta}_j) \prod_{i : \delta_i=0,I_{ij}=1}f{_j}(z_i \mid \boldsymbol{\theta}_j)\mathcal{I}(z_i \geq y_i^{obs}) \right], 
 \end{eqnarray}
 \noindent where $\boldsymbol{\theta}_j= (\eta_j, \boldsymbol{\beta}_j, \sigma^2_j)$,  $n_j= \sum_i I_{ij}$, $\sum_{j=1}^{k} n_j=n$ and $\mathcal{N}_j \sim f_j( \cdot \mid \boldsymbol{\theta}_j)$. Following Bayes’ theorem, the posterior distribution of the model parameters and latent variables, given the complete data $\mathbf{y} = (\mathbf{y}_1, \ldots, \mathbf{y}_{K})'$, is proportional to 
 \begin{small}
\begin{eqnarray}\label{eq16}\nonumber
 p\left(\boldsymbol{\eta}, \boldsymbol{\beta}, \sigma^2 \mid \mathbf{y}, \boldsymbol{\delta} \right) &\propto& f(\mathbf{y}^{obs}, \mathbf{z}, \boldsymbol{\delta} \mid \boldsymbol{\theta}) p(I \mid \boldsymbol{\eta}) \pi(\boldsymbol{\eta}, \boldsymbol{\beta},\sigma^2)  \\
  &\propto& { \prod_{j=1}^{K} \eta_j^{n_j}} \left[\prod_{i : \delta_i=1, I_{ij}=1}f{_{j}}(y_i^{obs} \mid \boldsymbol{\theta}_j) \prod_{i : \delta_i=0,I_{ij}=1}f{_j}(z_i \mid \boldsymbol{\theta}_j)\mathcal{I}(z_i \geq y_i^{obs}) \right] \prod_{i=1}^{n} p(I_i \mid \boldsymbol{\eta})  \\ \nonumber
 &\times& \pi(\boldsymbol{\eta}, \boldsymbol{\beta},\sigma^2). \nonumber
 \end{eqnarray} 
 \end{small}
 The Bayesian mixture model is completed by the prior distribution specification. We assume independence in the prior distribution with $\boldsymbol{\eta} \sim Dirichlet(K, \alpha)$, where $\sum_{j=1}^{K} \eta_j = 1$ and $\alpha=(\alpha_1, \ldots, \alpha_K)$ is a vector of hyperparameters, such that $\alpha_j > 0$;   $\phi_j=\frac{1}{\sigma^2_j}  \sim Gamma(a_j,b_j)$, the regression coefficients,  $\boldsymbol{\beta}_j \sim \mathcal{N}_j(\boldsymbol{m}_j, \tau^2_j\boldsymbol{I}_p)$ and $P(I_{ij}=1)=\eta_j$, for $i=1, \ldots, n$ and $j=1, \ldots, K$.
 
 The resulting posterior distribution in equation (\ref{eq16}) does not have closed form and we appeal to Markov chain Monte Carlo methods to obtain samples from the posterior distribution.  In particular, posterior samples are obtained through a Gibbs sampler algorithm, where the Markov chain is constructed by considering the complete conditional
distribution of each hidden variable given the others and the observations. The scheme is presented in the following subsection.
 
\subsection{Computational scheme}\label{sec2.3}

Assuming $K$ groups, it is possible to consider the Gibbs sampler algorithm in order to overcome the numerical integration condition of the data augmentation techniques, resulting in a computationally efficient algorithm.

We consider the following Bayesian Gaussian mixture survival model with data augmentation.
\begin{eqnarray}\nonumber
 \mathbf{y}_j \mid I_i, \boldsymbol{\beta}_j, \phi_j &\sim& \mathcal{N}_j(x_i^T \boldsymbol{\beta}_j, \phi^{-1}_j) \\ \nonumber
 I_i \mid \boldsymbol{\eta} &\sim& Categorical(K, \boldsymbol{\eta}) \\ \nonumber
 \boldsymbol{\eta} &\sim& Dirichlet(\alpha_1, \ldots, \alpha_K) \\ \nonumber
 \boldsymbol{\beta}_j &\sim& \mathcal{N}_j(\boldsymbol{m}_j, \tau^2_j\boldsymbol{I}_p) \\ \nonumber
 \phi_j &\sim& Gamma(a_j,b_j) \\ 
 p(z_i) &\propto& \mathcal{I}\left\{ z_i \geq y_i\right\} \nonumber
\end{eqnarray}
Algorithm \ref{algo:b} shows the scheme to estimate the parameters via data augmentation with censored observations. Details involved in obtaining the full conditional distributions  can be seen in Appendix \ref{apA}.
\begin{algorithm}
  \KwIn {Initialize all parameters $\boldsymbol{\eta}^{(0)}$, $\boldsymbol{\beta}_j^{(0)}$, ${\phi}_j^{(0)}$ for all $j=1, \ldots,K$}
 Update $I_i$ sampling from $I_i^{(k+1)} \sim I_i \mid \mathbf{y}, \boldsymbol{\eta}^{(k)}, \boldsymbol{\beta}_j^{(k)}, {\phi}_j^{(k)}$. \;
 Update $\boldsymbol{\eta}$ sampling from $\boldsymbol{\eta}^{(k+1)} \sim  \boldsymbol{\eta} \mid \mathbf{y}, I_i^{(k+1)}$. \;
  \lIf{$\delta_i=0$ for $i=1, \ldots, n$}{consider the data augmentation and define a latent variable $\mathbf{z}$ as
  $$y_i^{obs} \mid z_i , \left\{I_{ij}^{(k+1)}=1\right\} \sim \mathcal{NT}_{(-\infty,z_i)}\left(x_{ij}'\boldsymbol{\beta}_j^{(k)}, \phi_j^{-1(k)}\right)$$}
  Update $\phi_{j}$ sampling from $\phi_j^{(k+1)} \sim \phi_j \mid \mathbf{y}, I_i^{(k+1)}$.  \;
  Update $\boldsymbol{\beta}_j$ sampling from $\boldsymbol{\beta}_{j}^{(k+1)} \sim \boldsymbol{\beta}_{j} \mid \mathbf{y}, I_i^{(k+1)}, \phi_{j}^{(k+1)}$. \;
  Order $\boldsymbol{\beta}_{j}^{(k+1)}$ and arrange $\boldsymbol{\eta}^{(k+1)}$ and $\phi_{j}^{(k+1)}$ accordingly. \;
  $k=k+1$. Go back to step {\bf 1} until convergence.
  \caption{Gibbs sampler for a finite Gaussian mixture survival model with data augmentation}
  \label{algo:b}
\end{algorithm}
\subsection{Point estimation via Expectation Maximization}\label{EM}




Optimization methods to obtain maximum likelihood estimates are less computationally expensive than Monte Carlo estimation because they depend uniquely on numerical convergence. In order to obtain point estimates efficiently, we  consider the maximization of log-likelihoods 
 for the mixture model. 

Consider $K$ mixture components. In the classical context, the mixture model without censored data is described by equations \eqref{eq:mixture2} and \eqref{eq:mixture3}, respectively. Furthermore, without considering censored data and latent indicators of the mixture components, the log-likelihood function to be maximized is given by
\begin{equation}
    l\left(\left\{\eta_j\right\}_{j=1}^{K}, \left\{\boldsymbol{\beta}_j\right\}_{j=1}^{K}, \left\{\sigma^2_j\right\}_{j=1}^{K}\right) = \sum_{i=1}^{n}log \sum_{j=1}^{K} \eta_j f_j(Y_i \mid \boldsymbol{\beta}_j, \sigma^2_j).\end{equation}
 Notice that the expression depends on logarithms of sums, which cannot be simplified through logarithmic properties. The estimation, in this context, is exhaustive and without analytic or recursive forms for the maximum likelihood estimators of the model parameters.

In the mixture distribution context, it is very common to use the Expectation-Maximization algorithm proposed by \cite{ref_EM} which is an iterative mechanism to calculate the maximum likelihood estimator (MLE) in the presence of missing observations. Given the use of latent variables, and conditional on $I$,  equation \eqref{eq:mixture2} is valid. Thus, it provides a probability distribution over the latent variables together with a point estimate for parameters. If a prior distribution is assumed for the parameters, the joint posterior mode is obtained by the method.

Besides that, when we take into account the censored observed data, the data augmentation technique, as previously seen,  can be applied by including a new latent variable vector ${\bf z}$. According to  our model, censored observations are originated from a truncated normal distribution.

Assume that observation $y_i$ is censored. That is, there is an unobserved datum $z_i$ such that $z_i \mid \left\{I_{ij}=1 \right\} \sim f_j$ and $y_i^{obs} \mid z_i , \left\{I_{ij}=1\right\} \sim \mathcal{NT}_{(-\infty,z_i)}(x_{ij}' \boldsymbol{\beta}_j, \sigma^2_j)$. The strategy that we will adopt in the algorithm is to remove the truncation from the observed data $y_i^{obs}$ to obtain $z_i$, at each iteration $(k)$,   so that $y_{1:n}^{(k)} = (y_1^{obs}, y_2^{obs}, \ldots, y_{h}^{obs}, z_{h+1}^{(k)}, \ldots, z_{n}^{(k)})$, as previously seen in section \ref{sec2.2}.
 
Algorithm \ref{algo:c} is adapted for this context. For more details see \cite{ref_EM_censura}. Notice that $E(z_i \mid y_i^{obs}, \left\{I_{ij}=1\right\})$ and $Var(z_i \mid y_i^{obs}, \left\{I_{ij}=1\right\})$ denotes the expected value and  variance of a truncated Gaussian distribution, respectively. Although the computational cost (to obtain a point estimate) is smaller when compared to the proposal in section \ref{sec2.2} and \ref{sec2.3}, a disadvantage of this approach is that the EM algorithm is quite sensitive to the choice of initial parameters and does not take into account the uncertainty associated to parameter estimates. Besides, the EM algorithm will converge very slowly if a poor choice of initial value $\boldsymbol{\eta}^{(0)}$, $\boldsymbol{\beta}_j^{(0)}$, ${\sigma_j^{2(0)}}$ is selected.
\begin{algorithm}
 \KwIn {Initialize all parameters $\boldsymbol{\eta}^{(0)}$, $\boldsymbol{\beta}_j^{(0)}$, ${\sigma}_j^{2(0)}$ for all $j=1, \ldots,K$}
 
 {Define the latent variable $z$ as}
\[ z_i^{(k)} =
  \begin{cases}
   y_i^{obs},       &  \text{if} \quad \delta_i=1 \hspace{0.2cm} \forall i=1, \ldots, h\\
   \sum_{j=1}^{k} w_{ij}^{(k-1)}E(z_i \mid y_i^{obs}, I_i),  & \text{if} \quad \delta_i=0 \hspace{0.2cm} \forall i=h+1, \ldots, n.
  \end{cases}
\]  \;
 {Compute $w_{ij}$, the posterior probability that the obervation was generated from the  $j$-th mixture component}
 $$w_{ij}^{(k+1)} = \frac{\hat{\eta}_j^{(k)} f_j(z_i^{(k)} \mid \boldsymbol{\beta}_j^{(k)}, \sigma_j^{(k)})}{\sum_{j=1}^{K}\hat{\eta}_j^{(k)} f_j(z_i^{(k)} \mid \boldsymbol{\beta}_j^{(k)}, \sigma_j^{(k)})}, $$
 \noindent with $i=1, \ldots, n$ and $j=1, \ldots, K$. \;
 
 Update $\eta_j$ as $\hat{\eta}_j^{(k+1)}= \frac{1}{n}\sum_{i=1}^{n} w_{ij}^{(k+1)}$. \;
Update $\boldsymbol{\beta}_j$ and $\sigma^2_j$ as \;
 \begin{itemize}
     \item [-]  $\hat{\boldsymbol{\beta}}_j^{(k+1)}= \left(X'W_j^{(k+1)} X \right)^{-1} X'W_j^{(k+1)} \mathbf{z}^{(k)}$, $j=1, \ldots, K$ and $W_j^{(k+1)}=diag\left( \left\{ w_{ij}^{(k+1)}\right\}_{i=1}^{n}\right)$
     \item [-] $\hat{\sigma}^2{}^{(k+1)}_j=\frac{\sum_{i=1}^{n}w_{ij}^{(k+1)}\left(z_i^{(k)} - \mu_{ij}^{(k+1)}\right)^2 + \sum_{\left\{i : \delta_i=0\right\}} w_{ij}^{(k+1)}Var\left(z_i^{(k)} \mid y_i^{obs}, \left\{I_{ij}
     =1\right\}\right)}{\sum_{i=1}^{n}w_{ij}^{(k+1)}}$,
     \noindent where $\mu_{ij}^{(k+1)}= x_{ij}^{T}\boldsymbol{\beta}_j^{(k+1)}$.
 \end{itemize}
Run until convergence is achieved.
  \caption{Expectation Maximization algorithm for a finite Gaussian mixture survival model with data augmentation.}
  \label{algo:c}
\end{algorithm}


\newpage
\section{Applications}\label{sec3}
This section presents one simulated data set study to evaluate the performance and computational cost of our proposed model and one realistic simulated data set emulating a real portfolio considering lapse risks. 

\subsection{Simulated data set}\label{sec3.1}
In this subsection, we  return to the illustrative data set seen in section \ref{sec1.2}. Our aim is to compare the usual and  mixture survival log-normal models under a Bayesian approach through our proposals described in subsections \ref{sec2.1} and \ref{sec2.2}.

We simulate three scenarios: (i) a data set with  10\% of censored data; (ii) a data set with  40\% of censored data and (iii) a data set with 60\% of censored data, considering a mixture of $K=2$ components, with $\eta_1=\eta=0.6$. We would like to assess whether our proposal is efficient in sampling from the posterior distribution as well as its computational efficiency, for the model of interest. In addition, we vary the sample size ($n=1,000; 10,000; 50,000 ; 100,000$) in order to evaluate the computational cost through: (a) our proposal with data augmentation with censored data; (b) without data augmentation via \texttt{RStan} package available in \texttt{R} (\cite{standev2018rstan}, \cite{carpenter2017stan}), that is,  considering the survival likelihood given by equation (\ref{eq12}). Stan is a \texttt{C++} library for Bayesian modelling and inference that primarily uses the No-U-Turn sampler (NUTS) (see \cite{hoffman2014no}) to obtain posterior simulations given a user-specified model and data.

The survival time can be analysed according to $log(T_i)= \beta_0 + \beta_1 x_i + \varepsilon_i$, with $x$ the covariate that takes values ($x=0$ or $x=1$) and error $\varepsilon_i \sim N(0, \phi^{-1})$. We assign vague independent priors to the 
parameters in $\boldsymbol{\theta}$ with $\phi_j \sim Gamma(0.01,0.01)$, $\boldsymbol{\beta}_j \sim \mathcal{N}_j(\boldsymbol{0}, 100\boldsymbol{I}_2)$ and $\boldsymbol{\eta}\sim Dirichilet(\alpha_1=2, \alpha_2=2)$, for $j=1,2$. We run an MCMC chain for 20,000 iterations and consider the first 10,000 out as burn-in. The burn-in and lag for spacing of the chain were selected so that the effective sample size were around 1,000 samples.

\begin{figure}[H]
\centering
\begin{tabular}{cc}
   \includegraphics[width=5cm]{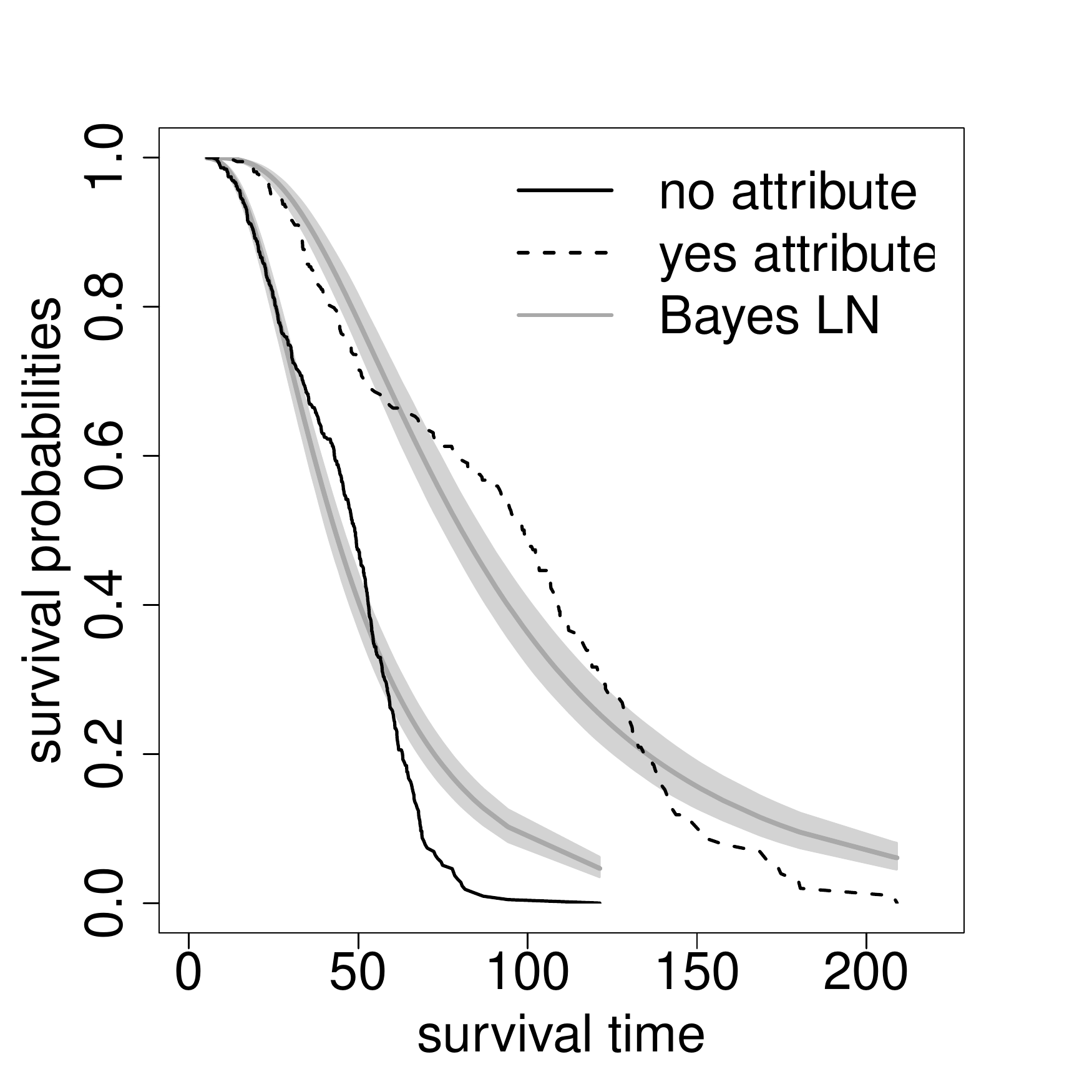} &    \includegraphics[width=5cm]{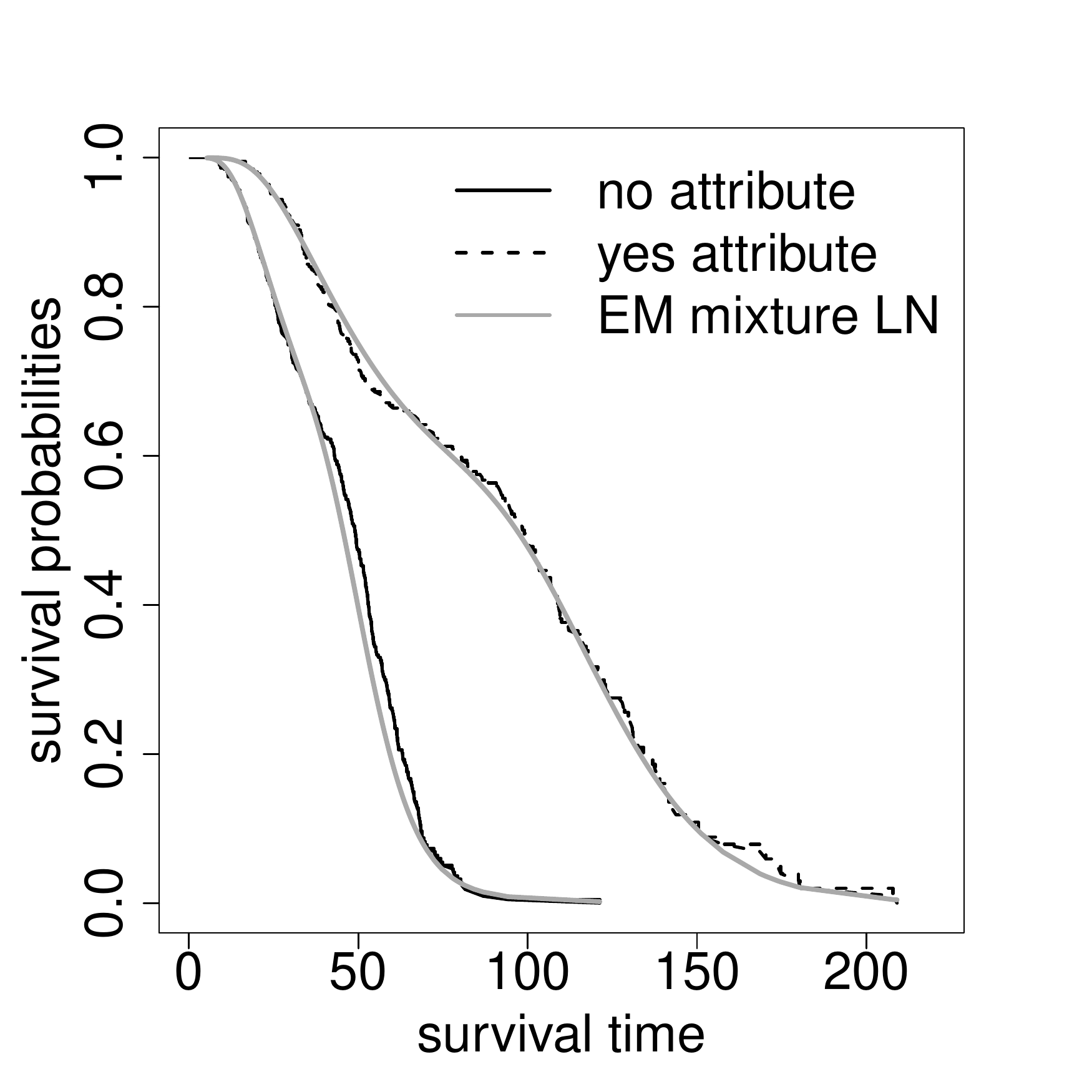}\\
   (a) Bayesian log-normal model & (b)  point estimation mixture log-normal model  \\
   \includegraphics[width=5cm]{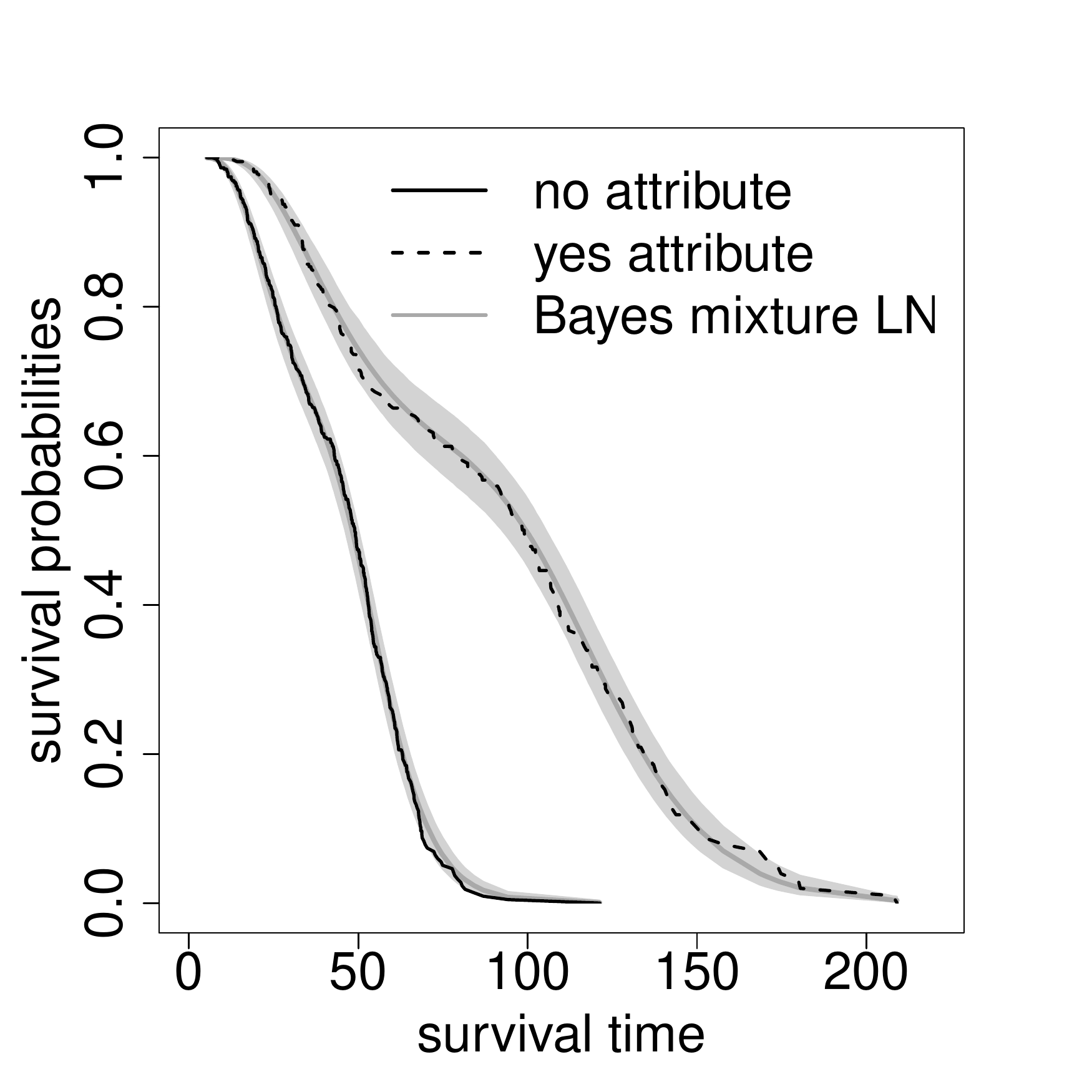}
   &   \includegraphics[width=5cm]{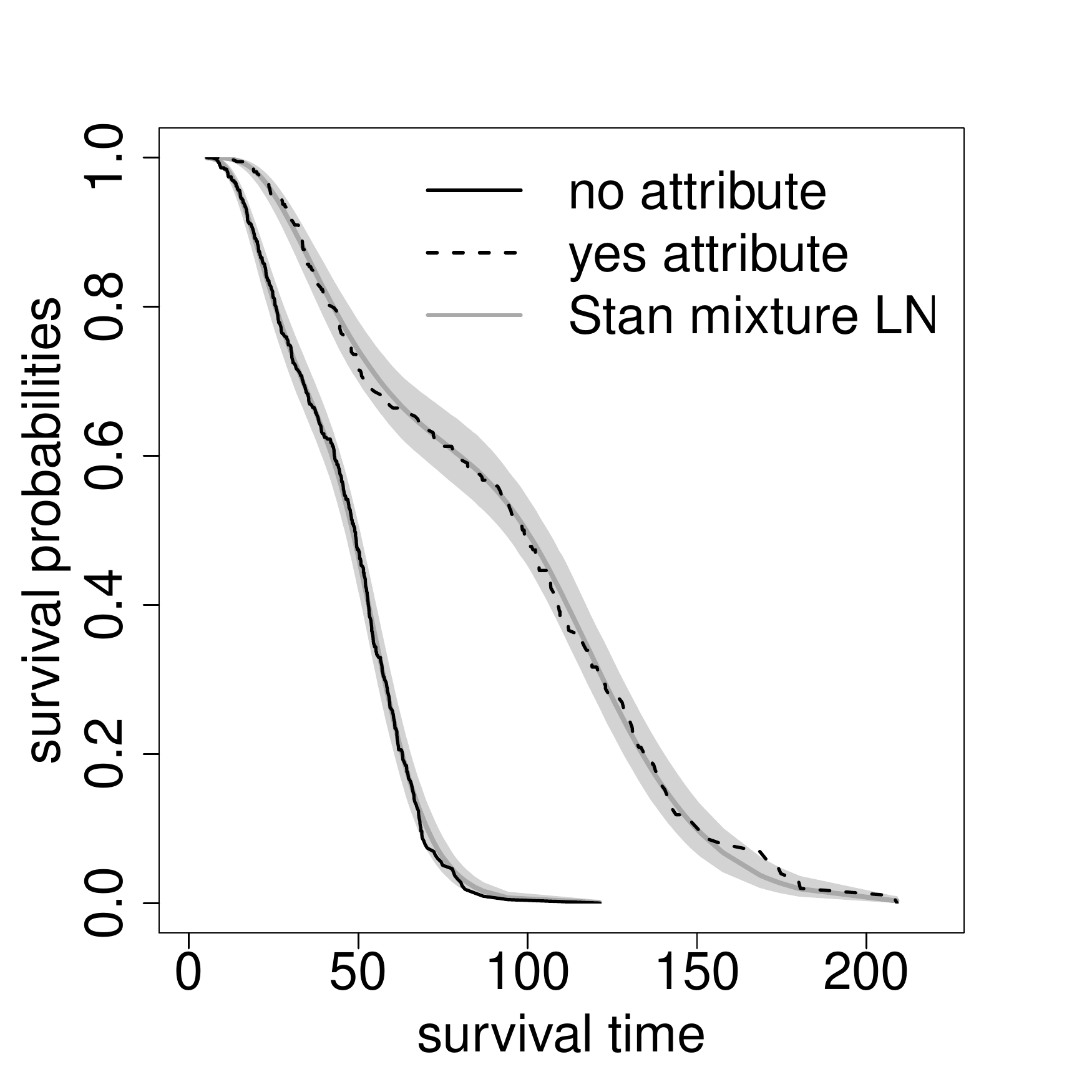}\\
   (c) Bayesian mixture log-normal model & (d) Stan Bayesian mixture log-normal model 
\end{tabular}
\caption{Simulated with 40\% censored data set: posterior survival probabilities with mean (grey line) and limits of 95\% credible interval: (a) the Bayesian Log-Normal model (BLN),  (b) the point estimation mixture Log-Normal model via EM algorithm (EMMLN), (c) the Bayesian mixture Log-Normal model (BMLN) with data augmentation, and (d) the Stan Bayesian mixture Log-Normal model (SBMLN), considering $n= 1,000$ policies.} \label{fig2}
\end{figure}

Figure \ref{fig2} illustrates the fit of the survival curves by the competing models considering a sample with 1,000 policies and 40\% rate of censorship (as seen in Figure \ref{fig1} (a)), the usual Bayesian log-normal model (without mixture, like in Figure \ref{fig1} (d)) and the Bayesian mixture Log-Normal model (see panels (c) and (d) in Figure \ref{fig2}). As can be seen, the proposed mixture model is able to accommodate different behaviours in the survival curves when compared to the usual Log-Normal model. In addition, the uncertainty associated with estimates is lower for our proposed mixture model. Panel (b), in Figure \ref{fig2}, exhibits the point estimation via Expectation-Maximization for the log-normal mixture modelling. The estimated survival curves via the EM algorithm follow the behaviour of the empirical Kaplan-Meier curves. Point estimates of the parameters of interest are reasonable compared to those obtained via Gibbs sampler techniques. See more details about the simulated data set in Appendix \ref{apB}.

Table \ref{tab:tab1} shows the posterior summaries for the survival Log-Normal model without mixture (Bayes LN) and considering Log-Normal mixtures via our proposal (Bayes Mixture LN, BMLN) and via Stan (Stan Bayesian Mixture LN, SBMLN), respectively. As already mentioned, the non-mixture model is not able to capture the behaviour of the survival curve. The structure of the non-mixture model does not allow the incorporation of mixture components in the coefficient estimates. On the other hand, the mixture Log-Normal model is capable of producing suitable estimates for the true parameters. Although the data augmentation proposal and the Stan method lead to similar point and interval estimates, the processing computational cost via Stan is much higher for all scenarios, as can be seen in Table \ref{tab:tab2}.  For the EM algorithm, 58 iterations were required until the parameters converged, which resulted in a computational time of 3.32 seconds. However, as already stated, the EM algorithm does not generate uncertainty measures associated with estimates. 


In this way, the use of Stan for large samples, high censored observations, and considering more covariates in the survival model can be prohibitive.
\begin{table}[H]
    \centering
    \caption{Posterior summaries comparison: mean and 95\% credibility for Bayes LN, Stan Mixture LN and Bayes Mixture LN; point estimation via EM Mixture LN, for a simulated data set with 40\% censorship rate and considering $n=1,000$ policies.}
    \begin{tabular}{|l|cccccc c|}
\hline \hline 
        &  \multicolumn{2}{c}{Bayes LN} &  \multicolumn{2}{c}{Stan Mixture LN} &  \multicolumn{2}{c}{Bayes Mixture LN} & EM Mixture LN  \\
    {\bf true}          &  mean & IC 95\% & mean & IC 95\% & mean & IC 95\% & pointwise \\
             \hline
 $\beta_{0,j=1}= 3.3 $  &3.76& (3.70,3.82) &  3.30    & (3.17,3.44) & 3.30&  (3.16,3.44) &   3.39\\
  $\beta_{0,j=2}=4.0 $  &-& - &   4.05  & (4.01,4.09) & 4.05& (4.02,4.08) & 3.98\\
   $\beta_{1,j=1}= 0.5$   &0.62& (0.53,0.72)  & 0.51    & (0.39,0.63)  & 0.51& (0.38,0.64) &  0.51 \\
    $\beta_{1,j=2}=0.8$   &-& - &   0.77  & (0.71,0.83)  & 0.77&  (0.72,0.82)  &0.84 \\
     $\sigma^2_{j=1}=0.3$   &0.61& (0.58,0.65)  &  0.23    & (0.17,0.31)  &0.24& (0.17,0.31)& 0.28 \\
     $\sigma^2_{j=2}=0.039$   &-& - &   0.04  & (0.03,0.06)  &0.04 & (0.03,0.06) & 0.04 \\
     $\eta = 0.60$ &-&- &  0.56   & (0.47,0.63) & 0.56 & (0.47,0.63)& 0.53\\
    \hline
    \end{tabular}
    \label{tab:tab1}
\end{table}

\begin{table}[H]
 \caption{Comparison of computational times (in seconds) involved in the Bayesian competing methods for simulated data sets  with 10\%, 40\% and 60\% rates of censorship and considering $n$ (size) policies. }
    \label{tab:tab2}
    \centering
    \begin{tabular}{|c|l| c |c|} 
    \hline \hline
{\bf \% censored} &  {\bf size $n$} & {\bf Bayes Mixture LN} & {\bf Stan Mixture LN} \\ 
    \hline
\multirow{4}{*}{10\%} &   1,000     &40.29    & 637.81  \\
 &   10,000     & 138.00  & 7,274.85   \\
  &  50,000     &532.80   &  23,913.10 \\
  &   100,000     & 1,157.4  &  65,452.20    \\
    \hline
\multirow{4}{*}{40\%}  &    1,000     & 42.90  & 831.70\\
  &  10,000     &  168.00 & 8,442.68  \\
  &  50,000     & 621.00  &  44,036.7  \\
  &   100,000     & 1,457.4 &  83,442.9  \\
    \hline
\multirow{4}{*}{60\%}  &    1,000     & 45.73  & 1,005.89 \\
  &  10,000     & 181.80 &    9,568.67\\
  &  50,000     & 1,048.8  & 49,751.10  \\
  &   100,000     & 1,816.2 & 106,977.0 \\
    \hline
    \end{tabular}
\end{table}

\subsection{A simulated data set in insurance}


In this subsection, We simulated a data set aiming to emulate the behaviour of a realistic portfolio in the private life insurance sector. We simulate a large data set emulating 100,000 policies over 100 months, taking into account heterogeneous lapse rates and realistic censored data with an approximate   $ 42.7\%$ censorship rate and two mixing components based on $\eta=0.6$, via the mixture Log-Normal model previously seen in Section \ref{sec2.1}. To illustrate, this data set contains individual policyholder information as well as information about the subscription.   

The factors considered in this study are: \texttt{gender} (male, female),  \texttt{age group} (18-29, 30-49, 60+),  \texttt{ policy type}  (standard, gold), where the level gold represents a segmentation of insureds that have a high insured capital and the premium \texttt{payment} mode (monthly, yearly, that is, regular premium or single premium) of the policy. Time to churn is the response variable of interest.

\begin{figure}[H]
\centering
\begin{tabular}{ccc}
   \includegraphics[width=5cm]{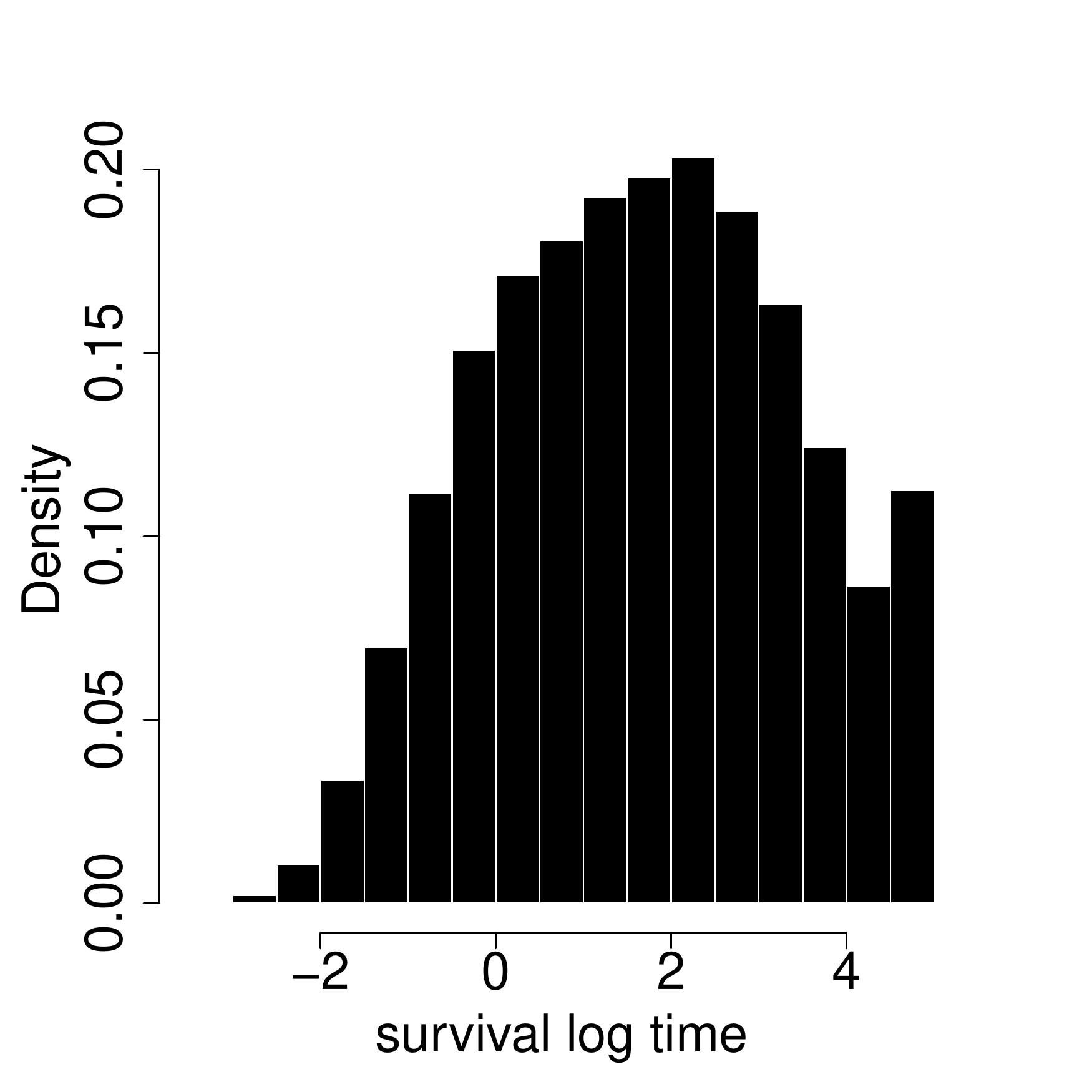} &    \includegraphics[width=5cm]{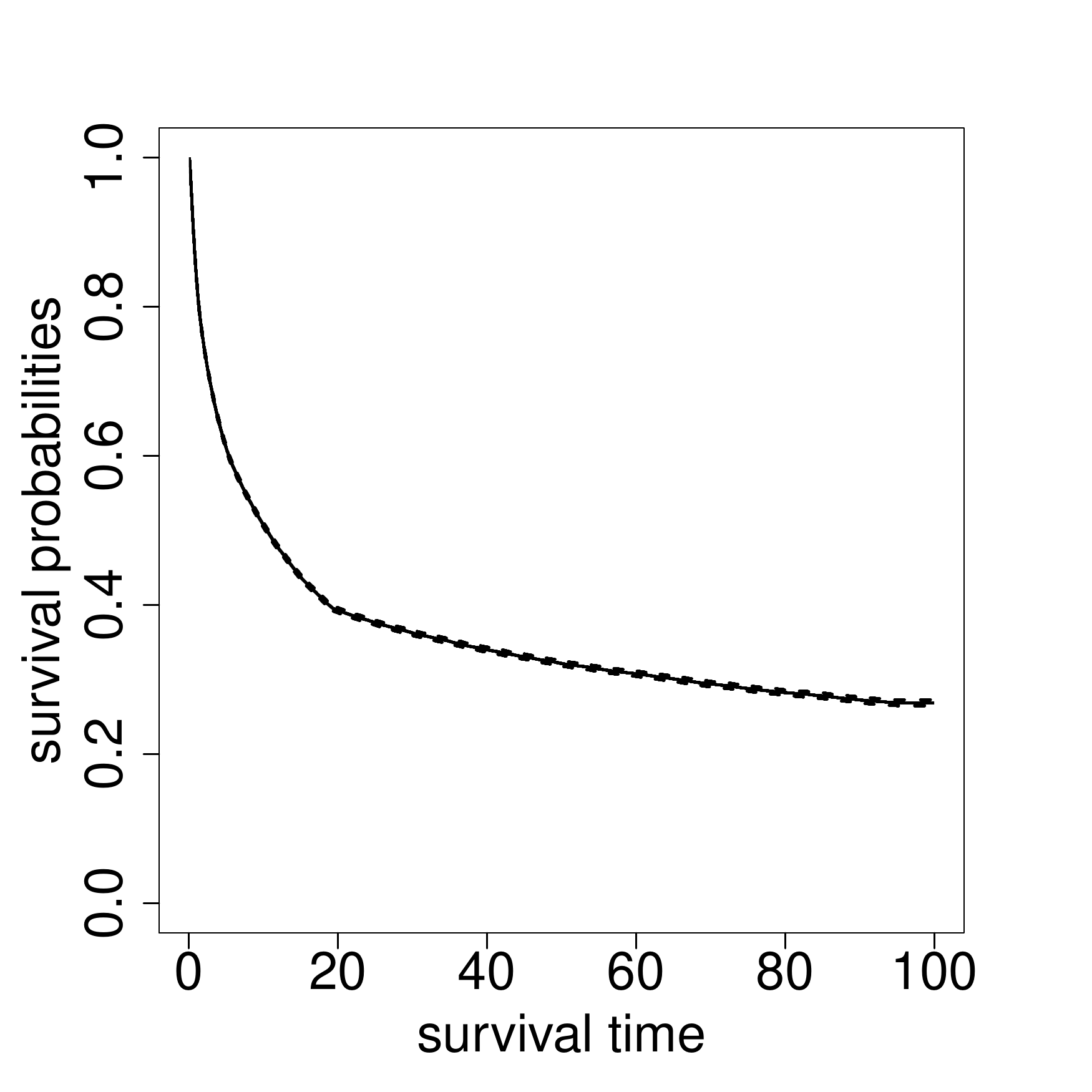}&
   \includegraphics[width=5cm]{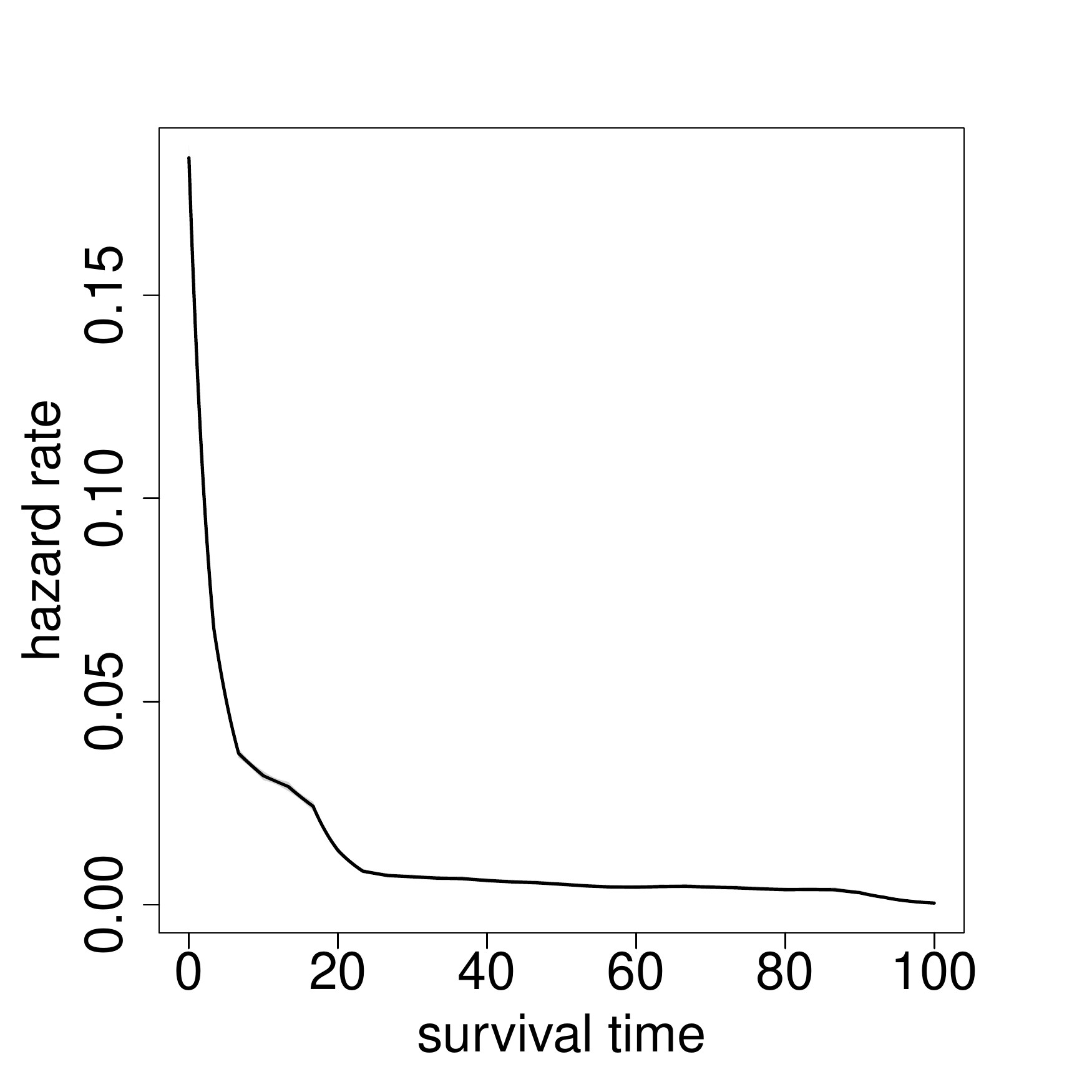}
  \\
  (a) survival log time & (b) empirical survival curve & (c) churn hazard curve \\
\end{tabular}
\caption{Simulated realist data set: a summary about the survival time with (a) survival time in log-scale, (b) empirical survival curve and (c) churn hazard curve.} \label{fig3}
\end{figure}

Panel (a) in Figure \ref{fig3} presents the simulated survival times in log-scale, indicating the mixing of two distributions. In panels (b)-(c), the empirical survival curve and hazard rate behaviour show that the lapse rate is higher and sharply falls for the first periods of time after subscription initiation, then it stabilises for some time, exhibits a peak close to 20 months, and then gradually decreases.

\begin{figure}[H]
\centering
\begin{tabular}{cc}
   \includegraphics[width=5cm]{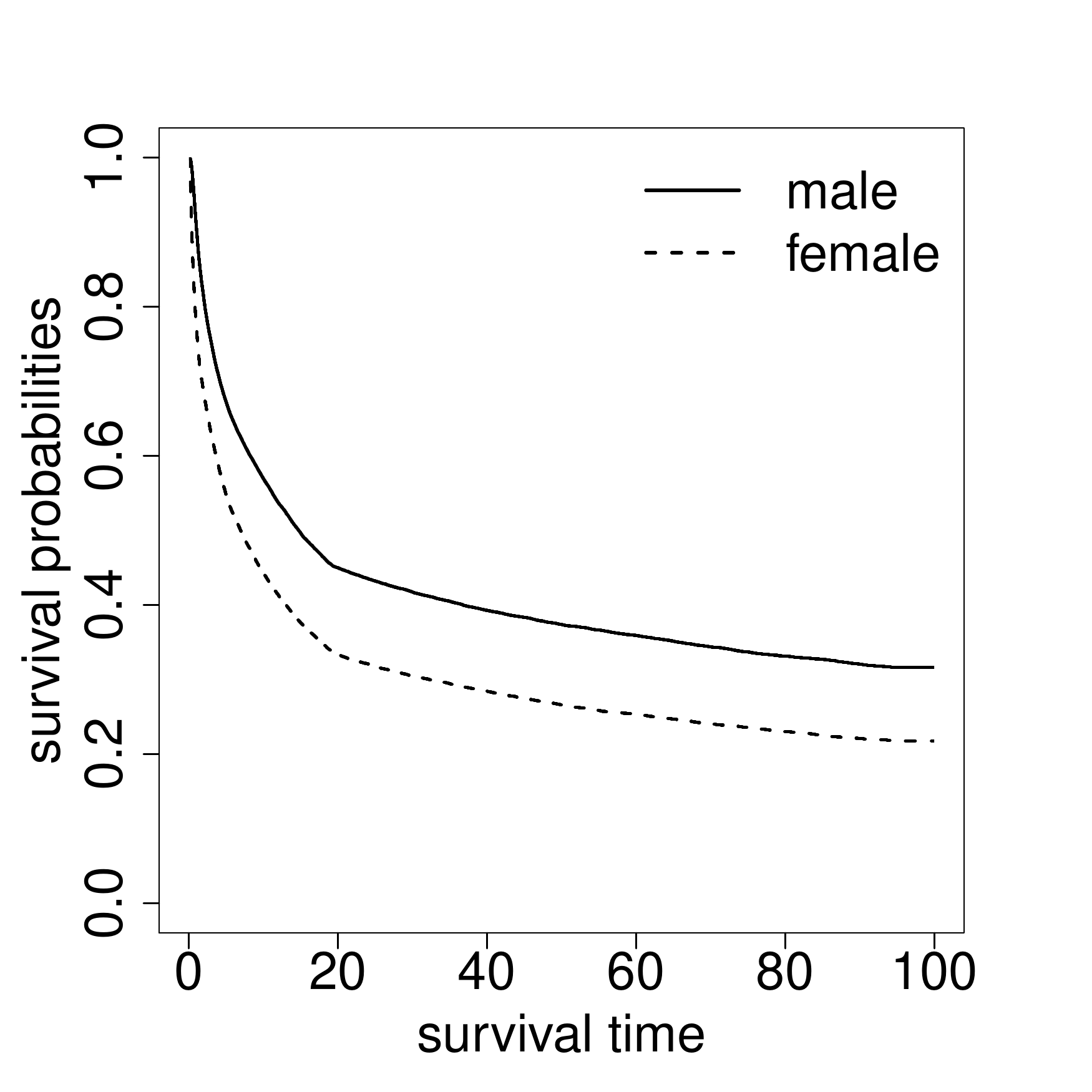} &    \includegraphics[width=5cm]{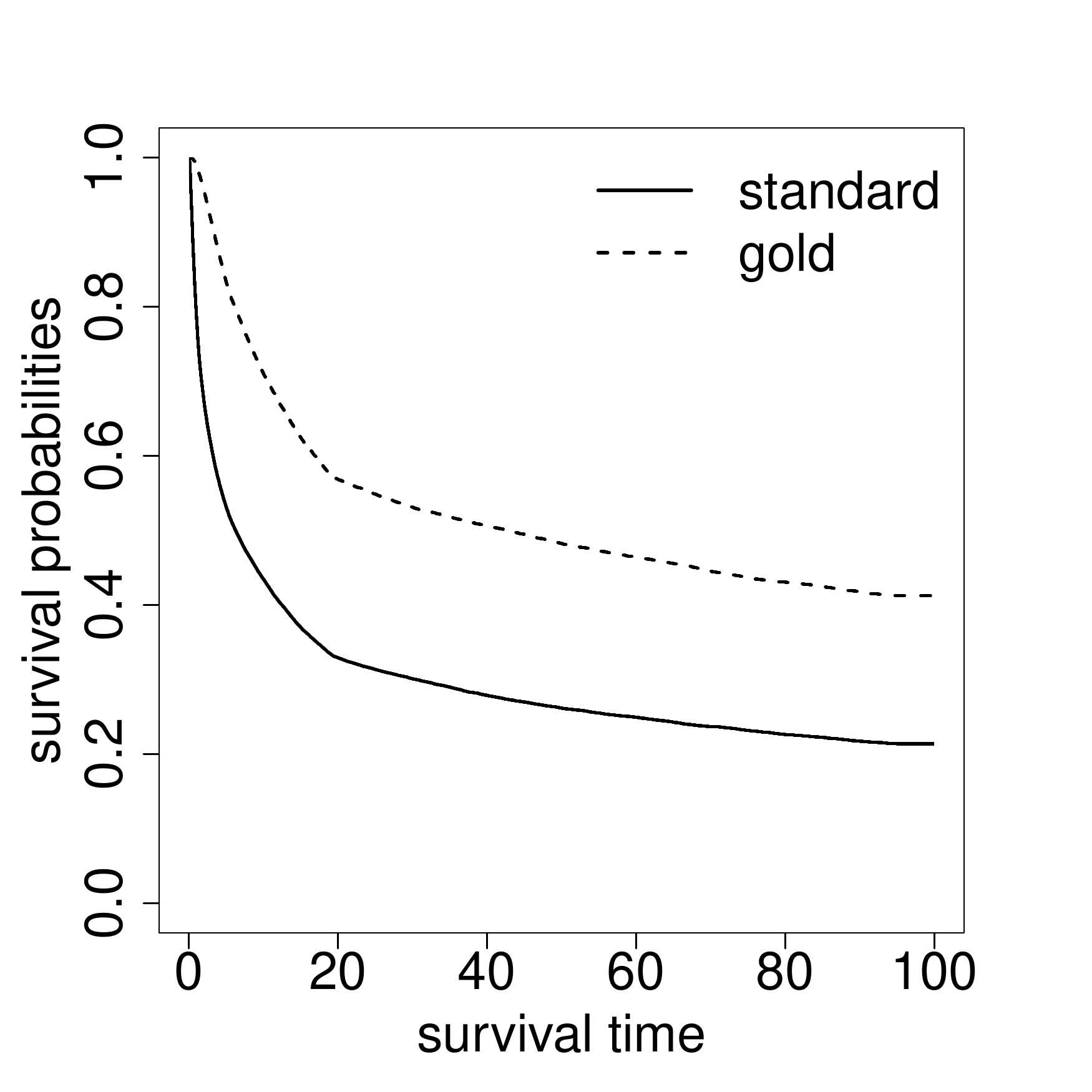}\\
   (a) gender  & (b) type policy \\ 
   \includegraphics[width=5cm]{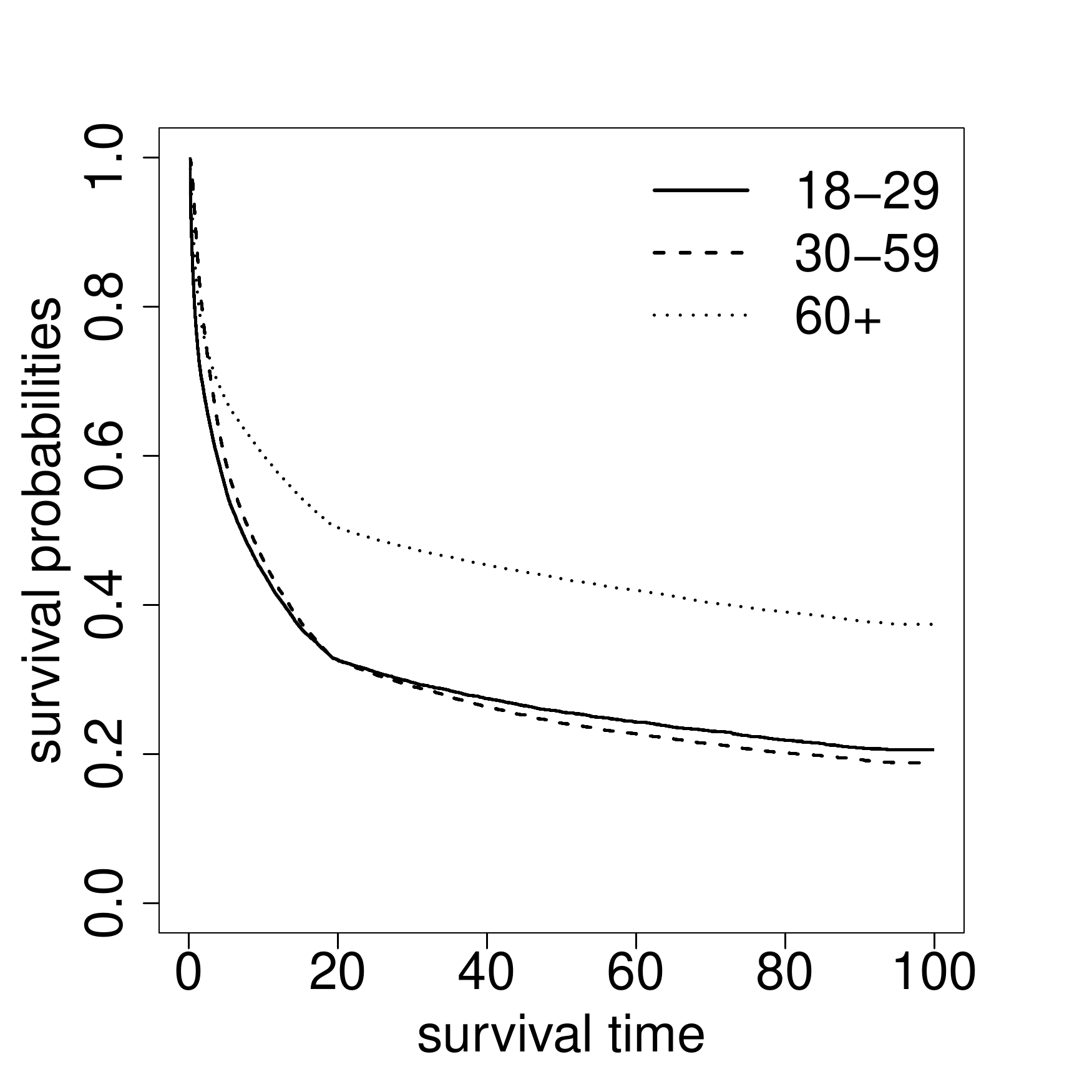} &
   \includegraphics[width=5cm]{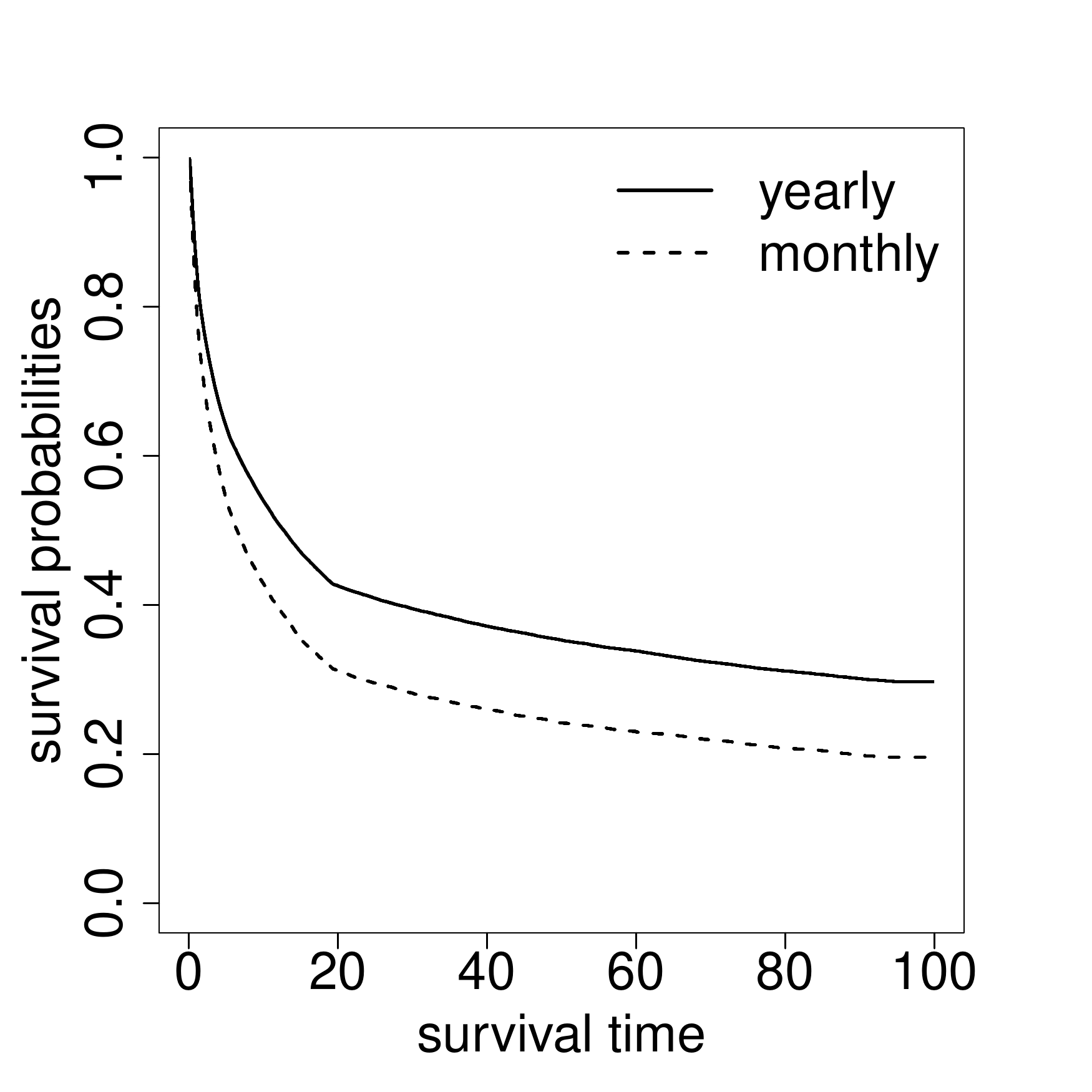}
  \\
  (c) insured age & (d) payment   \\
\end{tabular}
\caption{Marginal empirical survival curves via Kaplan-Meier for the covariates: (a) gender, (b) type policy, (c) age and (d) payment.} \label{fig4}
\end{figure}

Figure \ref{fig4} presents the marginal empirical Kaplan-Meier survival curve  for the variables of this study. As we can be seen, categories for each variable present particular behaviour and could be useful to understand the time to churn.  There is a noticeable drop in active insured in the first months after the policy subscription. Some reasons could be discussed, such as the policyholders who subscribe just to test a service or there may be some association with the premium payment mode.  For the age group , there is no visible difference in patterns of survival probability between the 18-29 and 30-59 group ages, except for a drop in during the first months of subscription and nearby the final portion of the survival curve. 

\begin{figure}[H]
\centering
\begin{tabular}{ccc}
   \includegraphics[width=5cm]{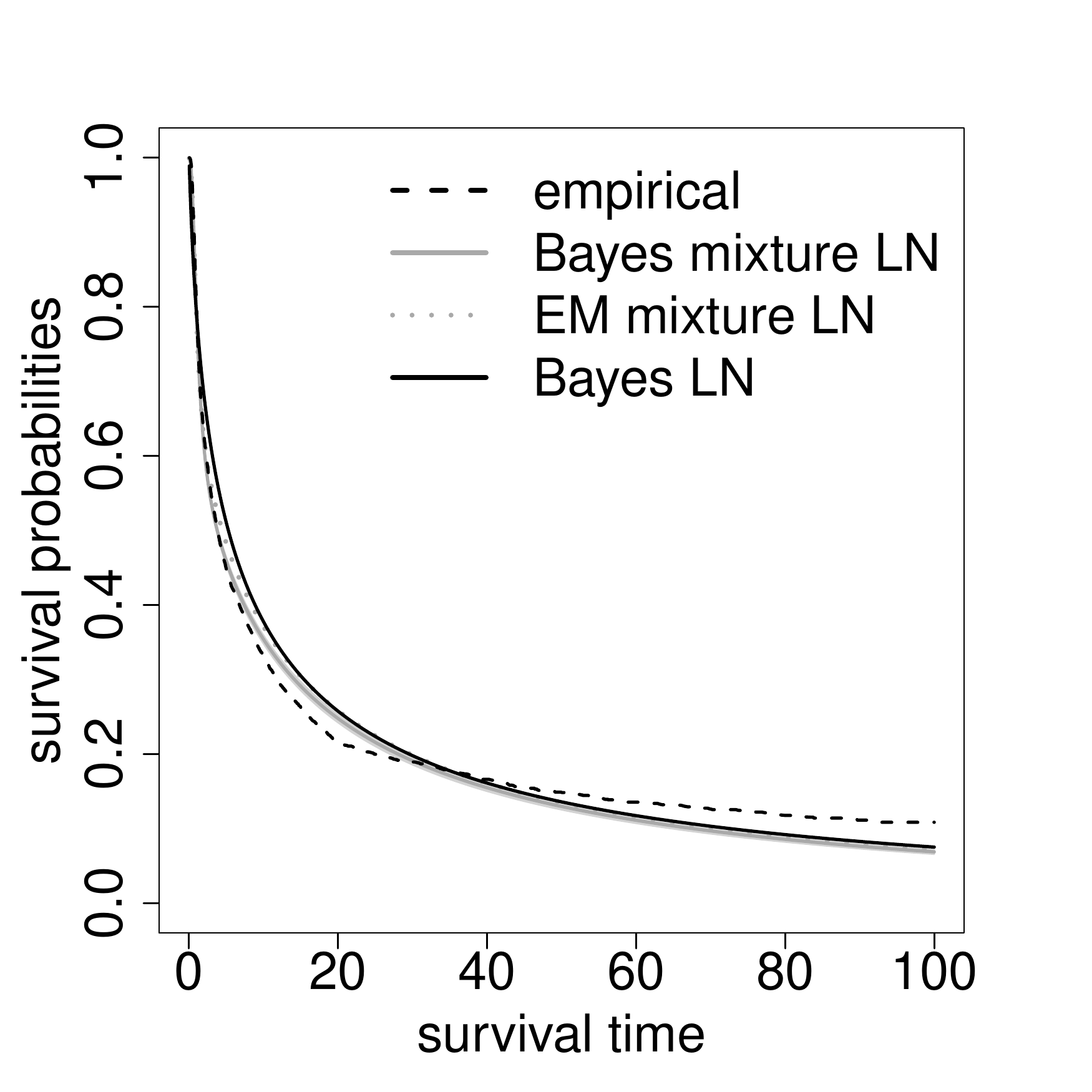} &   \includegraphics[width=5cm]{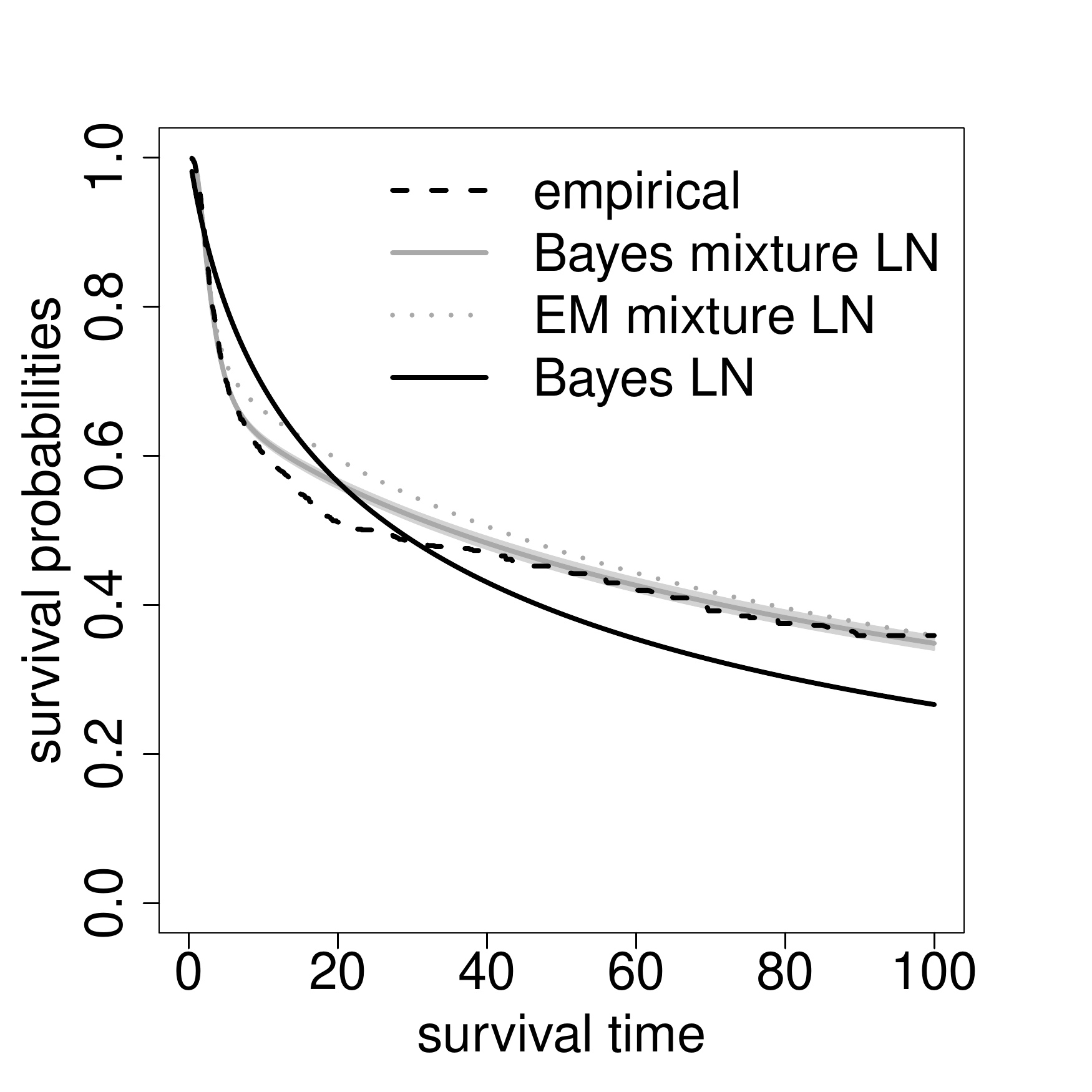} & \includegraphics[width=5cm]{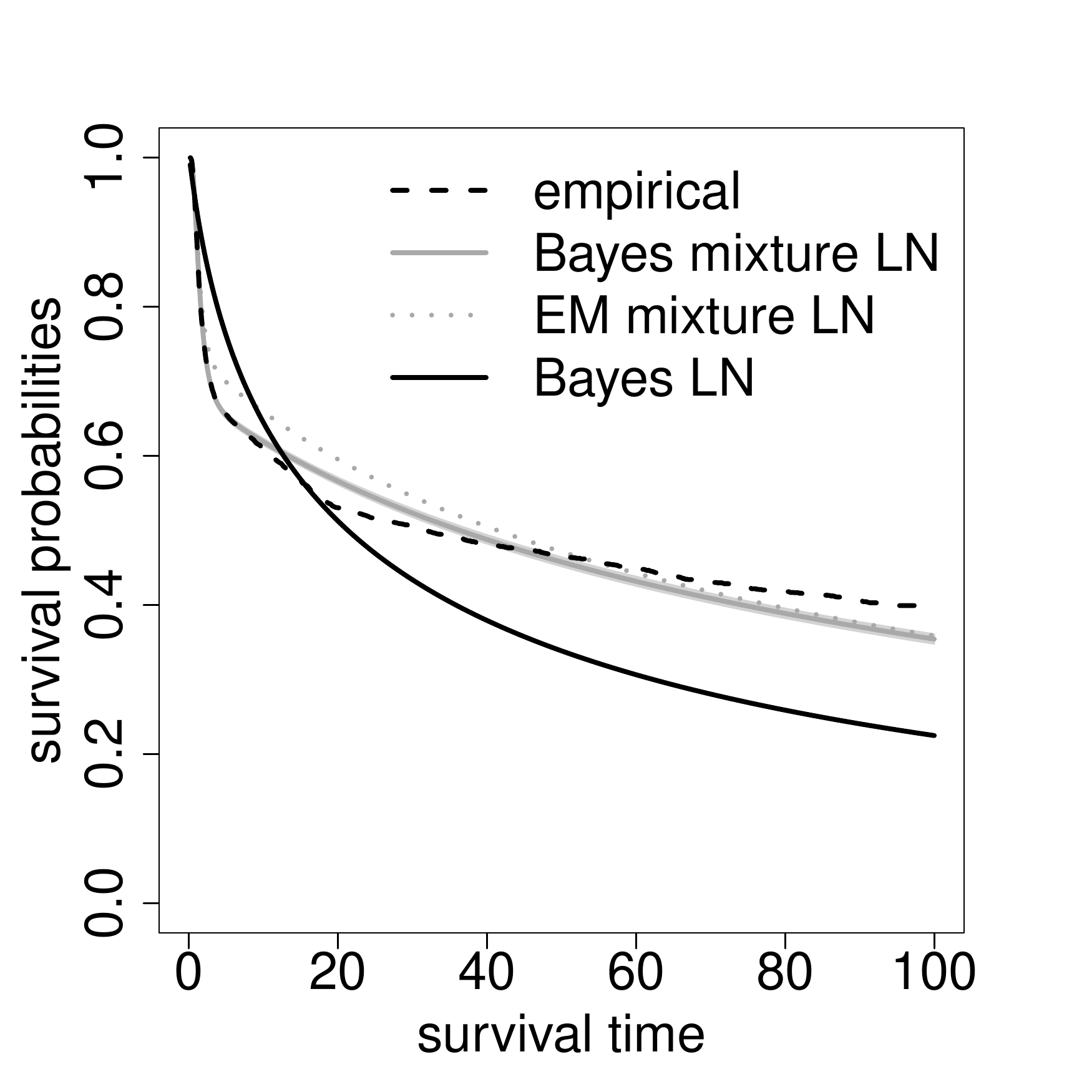}\\
   (a) scenario 1 & (b) scenario 2 & (c) scenario 3 \\
\end{tabular}
\caption{Summary survival curves: the empirical (dashed line), and the posterior for the Bayes mixture LN with 95\% credible interval (solid grey line), the EM mixture LN (dotted line) and the Bayes LN (solid black line).} \label{fig5}
\end{figure}

Figure \ref{fig5} shows the performance of the fitted curves for the competing models, the Log-Normal model and Log-Normal mixture model for some scenarios. 
Panel (a) represents the characteristics of the policyholder in scenario 1 (male, standard, 30-59 and monthly payment), panel (b) exhibits the fitted curves for scenario 2 (male, gold, 60+ and year), and panel (c) exhibits the estimated survival curves for scenario 3 (female, standard, 60+ and month). As we can see, policyholders in scenario 1 have lower persistence when compared to scenarios 2 and 3, respectively. This behaviour is understood due to the fact of the standard group and 30-59 age group preset survival empirical curves with more abrupt decay than 60+ and gold categories.

Our proposed Bayesian mixture survival model is able to capture the behaviour of the empirical curves (see Figure \ref{fig5}). Although the BMLN is reflects better the reality of this policyholders in terms of the survival curves, the EMMLN could be useful for a huge data set with million policyholders and return a better performance versus the usual survival model. Note that the BLN produces a poor fit due the fact this model is not allowed to access distinct behaviour of the curve at different times of the study. Besides that, the BMLN and EMMLN converge for the true values generated from the simulated data set. 


In order to understand the lapse of the policyholders, especially in the initial periods of the contract, we consider obtaining some probabilities to profile the policyholder to customer retention via the BMLN. We hope that the longer customers are with the insurance company, the less likely they are to cancel a policy. Table \ref{tab:tab3} shows a summary of the probabilities conditional on the hypothesis that the policyholder has survived the first three months, that is, conditional on the event $L=\left\{T > 3 \right\}$ for the scenarios in Figure \ref{fig5}. As one can see, the risk probability of a lapse occurring in the first year is higher for scenario 1 (0.394) when compared to scenarios 2 and 3 (0.247 and 0.123), respectively. On the other hand, the conditional probability that the policyholder maintains the contract term after 36 months, having survived to the first  months ($P(T \geq 36 \mid T> 3) $) is high for all considered scenarios. Other probabilities can be evaluated in order to understand the profile of the company's policyholders. To illustrate, Table \ref{tab:tab4} exhibits the results obtained conditioning on other events, such as $\left\{ T>12\right\}$, $\left\{T>24\right\}$ and $\left\{T>36\right\}$. As expected, the probability of churn decreases, having the insured persisted in the insurance company for long periods, that is, the lapses reduce substantially with increasing policy age. Furthermore, when identifying insured profiles, we are able to understand which profiles need more attention, thus the insurance company could develop retention  strategies focused on these policyholders.
\begin{table}[H]
\caption{Probability for the time to churn conditional on the event $L=\left\{T > 3 \right\}$ for the scenarios 1, 2 and 3.}\label{tab:tab3}
    \centering
    \begin{tabular}{|l | cccc|}
    \hline \hline
     {\bf profile}   &$P(T\leq 12\mid L)$&$P(12<T< 24 \mid L)$ & $P(24<T<36 \mid L)$ & $P(T \geq 36 \mid L)$   \\
     \hline
       scenario 1   & 0.394  & 0.194& 0.100& 0.312 \\
           scenario 2   &0.247  &0.078 & 0.059& 0.616 \\
               scenario 3   & 0.123 &0.086 &0.067 & 0.724 \\
               \hline
    \end{tabular}
\end{table}
\begin{table}[H]
\caption{Probability for the time to churn conditional on the events $\left\{ T>12\right\}$, $\left\{T>24\right\}$ and $\left\{T>36\right\}$ for the scenarios 1, 2 and 3.}\label{tab:tab4}
    \centering
    \begin{tabular}{|c|c c c| }
    \hline \hline
     {\bf profile}   & $P( T \leq 24 \mid T>12)$ & $P(T \leq 36 \mid T>24)$ & $P(T \leq 48 \mid T>36)$  \\
      \hline
    scenario 1     &   0.320 & 0.243 &   0.200\\
        scenario 2     &   0.104  & 0.087  &  0.077 \\
            scenario 3     &  0.098   & 0.085  & 0.076 \\
            \hline
    \end{tabular}
  \end{table}


\section{Conclusions}\label{sec4}
We have proposed a flexible survival mixture model that extends the usual survival model and accommodates different behaviour  of $T_i$ over time, via mixtures of parametric models and considering censored survival times. The proposal combines the mixture distributions based on  \cite{fruhwirth2006finite} and data augmentation techniques proposed by \cite{tanner1987calculation}. Full uncertainty quantification is available through the Bayesian distributions which are obtained via MCMC methods. Furthermore, we proposed an efficient sampling algorithm for inference summarised in point estimates, via Expectation-Maximization algorithm.

We performed extensive simulation studies to investigate the ability of the proposed model to capture different survival curves with the mixture survival proposal. Our simulated examples indicate that the data generating model (with mixture of distributions) provides the best fit and indicates that the usual survival models are not able to capture the change in behaviour for the churn times. Besides, our designed solution combining the Gibbs sampler and data augmentation is faster than the solution provided by the Stan package. 

We conclude that allowing for a flexible survival model allows for more realistic description of the behaviour of the survival probability curves, for the factors affecting times to an event such as surrender. 

\section*{Acknowledges}
We are grateful to LabMA/UFRJ (Laboratório de Matemática Aplicada of the Universidade Federal do Rio de Janeiro) based in Brazil for financial support and to its members for the very enriching discussions. In particular, we thank Dr. Mário M. C. de Oliveira and Dr. Bruno Costa for providing support in this research. We are also grateful for the research assistantship from the current postgraduate students Rafael Cabral Fernandez and Guilherme dos Santos in constructing the survival mixture algorithm and simulation data set and the former undergraduate student Bryan Suhett for assisting with the implementation of the EM algorithm via mixture.  

\appendix
\section{Appendix}\label{apA}
The prior distributions considered for the parameters, the complete conditional
distributions and proposal densities used in the MCMC algorithm are detailed as
follows. 

\subsection{Bayesian mixture survival model}
Consider the likelihood given in Equation (\ref{eq15}) and mixture components in Equation (\ref{eq:mixture2}).

The conditional distribution of each $I_i$, $i=1, \ldots, n$, given all other parameters and prior distribution $I_i \mid \boldsymbol{\eta} \sim Categorical(K, \boldsymbol{\eta})$, $\boldsymbol{\eta}=(\eta_1, \ldots, \eta_K)$ is given by
\begin{eqnarray} \nonumber
p(I_{ij}= 1 \mid y_i, \boldsymbol{\eta}, \cdot) &\propto& f(y_i \mid \left\{I_{ij}=1\right\}, \boldsymbol{\eta})\pi(I_{ij}= 1 \mid \boldsymbol{\eta}) \\ \nonumber
 &\propto& \frac{\eta_j f_j(y_i)} {\sum_{j=1}^{K} \eta_j f_j(y_i)}. \nonumber
\end{eqnarray}
If $y_i \mid \left\{I_{ij}=1\right\} \sim \mathcal{N}_j(\mu_{ij}, \phi_j^{-1})$, then
\begin{equation}
    p(\left\{I_{ij}=1\right\} \mid y_i, \boldsymbol{\eta}, \boldsymbol{\beta}_j, \phi_j) = \frac{\eta_j \mathcal{N}_j(y_i \mid \boldsymbol{\beta}_j, \phi_j)}{\sum_{j=1}^{K} \eta_j \mathcal{N}_j(y_i \mid \boldsymbol{\beta}_j, \phi_j)}.
\end{equation}
Thus, the marginal posterior distribution is $I \mid \mathbf{y} \sim Categorical\left(K;  \frac{\eta_1 \mathcal{N}_1(y_i \mid \boldsymbol{\beta}_1, \phi_1)}{\sum_{j=1}^{K} \eta_j \mathcal{N}_j(y_i \mid \boldsymbol{\beta}_j, \phi_j)}, \ldots, \frac{\eta_K \mathcal{N}_K(y_i \mid \boldsymbol{\beta}_K, \phi_K)}{\sum_{j=1}^{K} \eta_j \mathcal{N}_j(y_i \mid \boldsymbol{\beta}_j, \phi_j)} \right)$.

The conditional distribution for $\eta \mid I, \cdot$, considering a prior distribution as $\boldsymbol{\eta} \sim Dirichilet(K; \alpha_1, \ldots, \alpha_K)$ is given by
\begin{eqnarray}\nonumber
 p(\boldsymbol{\eta} \mid I, \cdot) &\propto& f(I \mid \boldsymbol{\eta}, \cdot)\pi(\boldsymbol{\eta}) = \prod_{i=1}^{K} p(I_i \mid \boldsymbol{\eta}, \cdot) \pi(\boldsymbol{\eta}) = \left[ \prod_{i=1}^{n} \prod_{j=1}^{K} \eta_j^{[I_{ij}=1]}\right] \prod_{j=1}^{K} \eta_j^{\alpha_j-1}\\ \nonumber
 &\propto&  \prod_{j=1}^{K}  \eta_j^{\alpha_j- 1+ \sum_{i: I_{ij}=1}[I_{ij}=1]} = \prod_{j=1}^{K} \eta_j^{(\alpha_j+ n_j)-1}, \nonumber
\end{eqnarray}
\noindent where $n_j= \sum_{i} I_{ij} $. Thus,  $\boldsymbol{\eta} \mid I, \cdot \sim Dirichilet(\alpha_1^{*}, \ldots, \alpha_K^{*})$, with $\alpha_j^{*}= n_j + \alpha_j$, $j=1, \ldots, K$.

For each group $j$, we can compute sample from each of components $\phi_j= 1/\sigma_j^2$ individually. For $\phi_j \sim Gamma(a_j, b_j)$, 
\begin{eqnarray}\nonumber
   p(\phi_j \mid \mathbf{y}_j, \cdot) &\propto& f(\mathbf{y}_j \mid \phi_j, I, \cdot)\pi(\phi_j) \\ \nonumber
    &\propto& (\phi_j)^{n_j/2} exp\left\{-\phi_j \left(\frac{(\mathbf{y}_j - \mathbf{x}_i^{T}\boldsymbol{\beta}_j)^{T}(\mathbf{y}_j - \mathbf{x}_i^{T}\boldsymbol{\beta}_j)}{2}\right)\right\}\phi_j^{a_j-1} \exp \left\{ -\phi_j {b_j}\right\} \\ \nonumber
    &\propto& \phi_j^{a_j + n_j/2 - 1} exp\left\{-\phi_j \left({b_j} + \frac{(\mathbf{y}_j - \mathbf{x}_i^{T}\boldsymbol{\beta}_j)^{T}(\mathbf{y}_j - \mathbf{x}_i^{T}\boldsymbol{\beta}_j)}{2}\right)\right\}.
\end{eqnarray}
The conditional distribution of $\phi_j \mid \mathbf{y}_j, \cdot \sim Gamma\left( {a_j} + \frac{n_j}{2} ; {b_j} + \frac{(\mathbf{y}_j - \mathbf{x}_i^{T}\boldsymbol{\beta}_j)^{T}(\mathbf{y}_j - \mathbf{x}_i^{T}\boldsymbol{\beta}_j)}{2} \right)$.

For $\boldsymbol{\beta}_j \sim \mathcal{N}_j(m_j, \tau^2_j)$, with $\boldsymbol{\beta}_j= (\beta_{0j}, \beta_{1j}, \ldots, \beta_{pj})$. The conditional distribution is given by,
\begin{eqnarray} \nonumber
  p(\boldsymbol{\beta}_j \mid \mathbf{y}_j, \cdot) &\propto& f(\mathbf{y}_j \mid \boldsymbol{\beta}_j, \cdot)\pi(\boldsymbol{\beta}_j) \\ \nonumber
\end{eqnarray}
For latent variable $Z$, when $\delta_i=0$, $i=1, \ldots, n$, we consider the data augmentation. Thus, $p(z_i \mid \left\{I_{ij}=1\right\}, \boldsymbol{\beta}_j, \phi_j, \cdot) \propto \mathcal{I}(y_i \geq y_i)$, we have:
\begin{eqnarray}\nonumber
 p(z_i \mid I\left\{I_{ij}=1\right\}, y_i, \cdot) &\propto& f(y_i \mid \left\{I_{ij}=1\right\}, \cdot)\pi(z_i \mid \left\{I_{ij}=1\right\}) \\ \nonumber
 &\propto& f(y_i^{obs}, z_i \mid \left\{I_{ij}=1\right\}, \cdot)\pi(z_i) \\ \nonumber
 &\propto& f(z_i \mid \left\{I_{ij}=1\right\})\pi(z_i)= \exp\left\{ - \frac{\phi_j}{2}(z_i - \mathbf{x}_i^{T}\boldsymbol{\beta}_j)^2\right\}\mathcal{I}(z_i \geq y_i). \nonumber
\end{eqnarray}
The conditional distribution of $z_i \mid \left\{I_{ij}=1\right\}, y_i \sim \mathcal{NT}(\mathbf{x}_i^{T}\boldsymbol{\beta}_j, \phi^{-1}_j)$.

\section{Some results for simulated data sets}\label{apB}

\subsection{Simulated data set}
\begin{figure}[H]
    \centering
    \begin{tabular}{cc}
    \includegraphics[width=5cm]{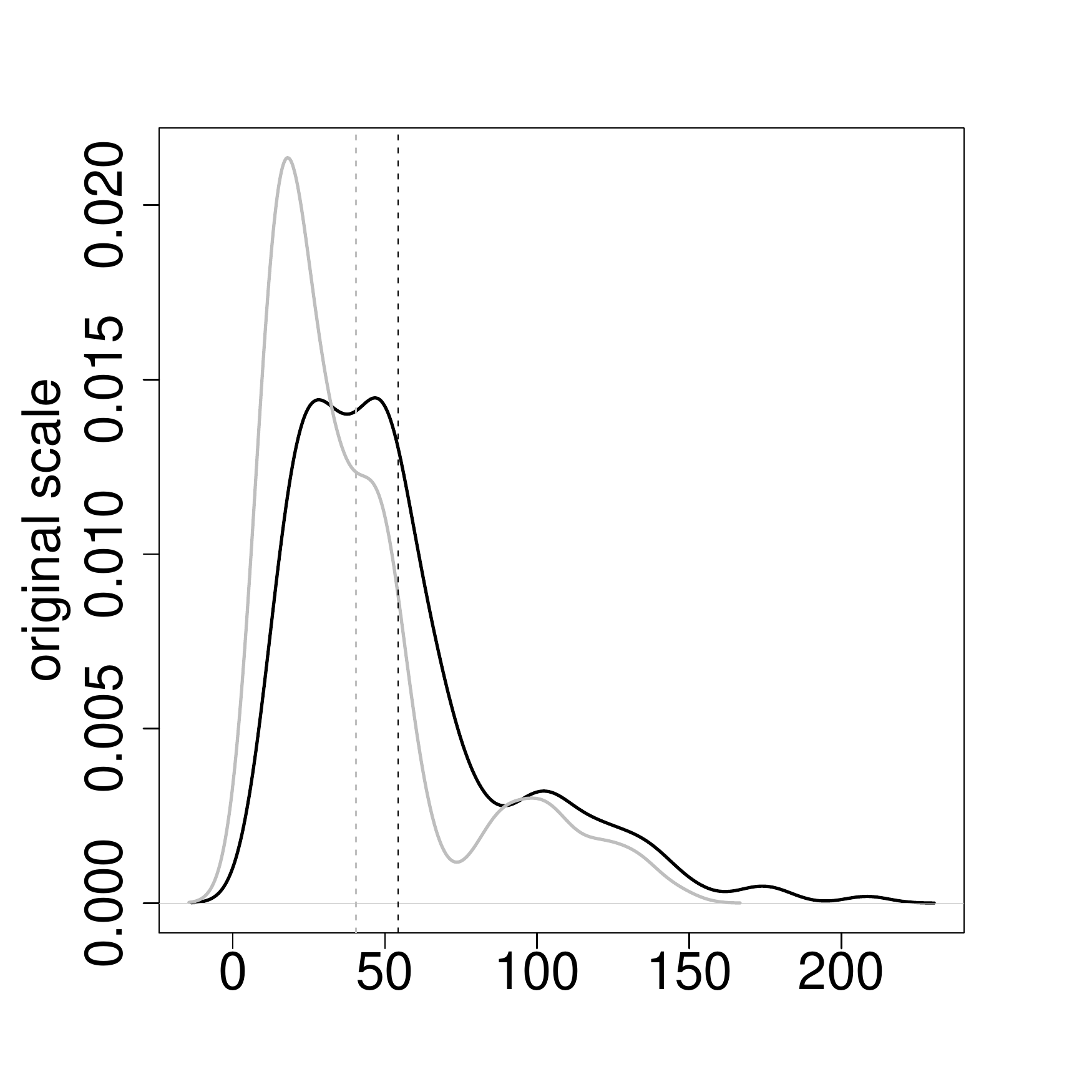} &
       \includegraphics[width=5cm]{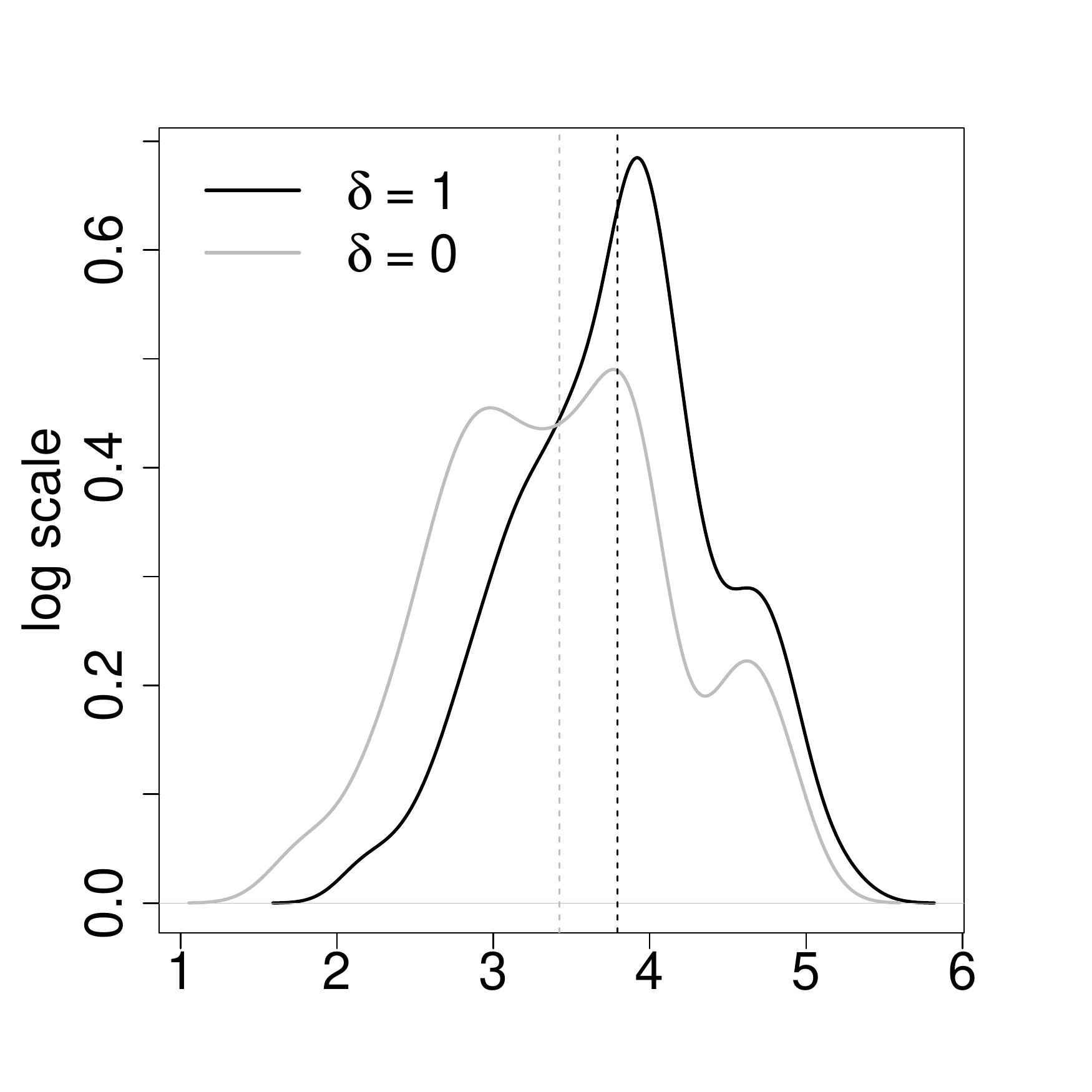}\\
       (a) original scale & (b) log scale \\
    \end{tabular}
     \caption{Simulated with 40\% censored data set considering $n=1,000$ policies: (a) original scale $T_i$ and (b) log-scale $log(T_i)$. Besides, 60\% come from $\mathcal{N}_1(\cdot, \cdot)$ and 40\% come from $\mathcal{N}_2(\cdot, \cdot)$. Dashed lines are the mean for $\delta=0$ and $\delta=1$, respectively.}
    \label{fig1apB}
\end{figure}

In this appendix we present the generation of observations via mixture distribution considering the presence of censored data in the sample in Section \ref{sec3.1}. The data set was simulated considering variable $Y_i= log(T_i)$ as the logarithmic the duration of a policy $i$ before termination from a mixture of Gaussian distribution with $K=2$ components given by 
\begin{equation}
    f(y_i \mid \boldsymbol{\theta}) = \eta \thinspace \mathcal{N}_1 (\mu_{i1},\sigma^2_1) + (1-\eta) \mathcal{N}_2(\mu_{i2}, \sigma^2_2), \thinspace \thinspace i=1,\ldots,n,
\end{equation}
\noindent where $\boldsymbol{\theta}=(\eta, \boldsymbol{\beta}, \sigma^2)$ the parametric vector of interest. The mean for $j=1,2$ is given by $\mu_{i,j=1}=\beta_{01} + \beta_{11} x = 3.3 + 0.5x$ and $\mu_{i,j=2}=\beta_{02} + \beta_{12} x = 4.0 + 0.8x$ and variance as $\sigma^2_1=0.3$ and $\sigma^2_2=0.039$, respectively. The weight $\eta_1=\eta$ is equal to 0.6 and $x$ represents the covariate that takes values ($x=0$, no attribute and $x=1$, yes attribute). In the presence of censored observation, that is, $\delta_i=0$, we generate the censored observation $y_i^c$ from a truncated Gaussian distribution as
\begin{equation}
    y_i^c \mid \cdot \sim \mathcal{NT}_{(-\infty, y_i]}\left(\mu_{ij}, \sigma^2_j\right), \forall \thinspace \thinspace j=1, \ldots, K.
\end{equation}
Thus, in inferential processes with simulated data, the likelihood will be based on the original observations $y_i$ for $\delta_i=1$ and the censored ones, $ y_i^c$, for $\delta_i=0$, emulating the observable information in real practical situations.
Figure \ref{fig1apB} shows the mixture distribution with 40\% censored data considering $n=1,000$ policies. Panels (a) show the behaviour in the original scale, that is, $T_i$ from a log-normal distribution and (b) in the log scale. See that the mixture proportions are indeed based on the probabilities that were defined, i.e., $\eta=0.6$ and $1-\eta=0.4$, respectively. 

\begin{figure}[H]
\begin{center}
    \begin{tabular}{cccc}
      \includegraphics[width=4cm]{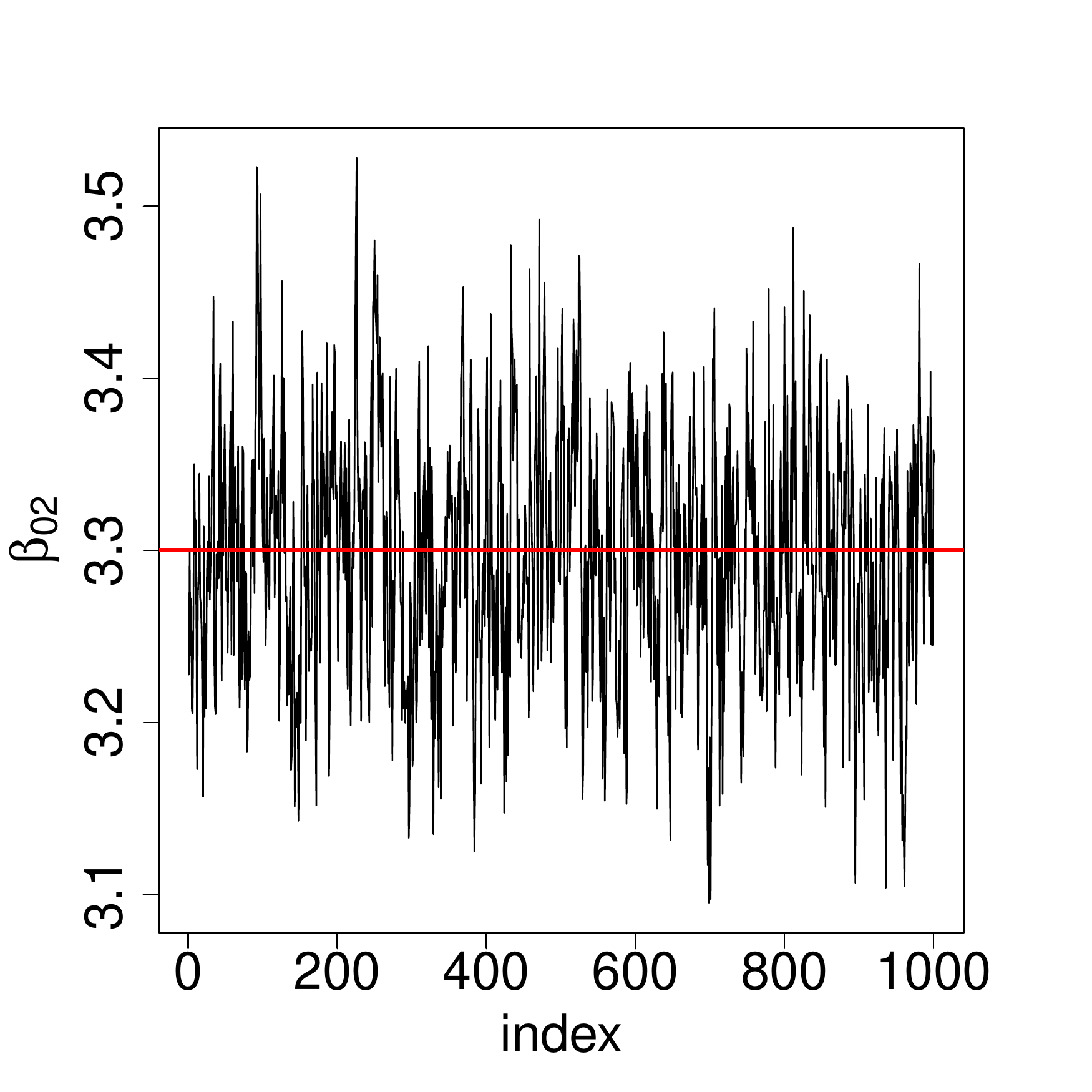} &
       \includegraphics[width=4cm]{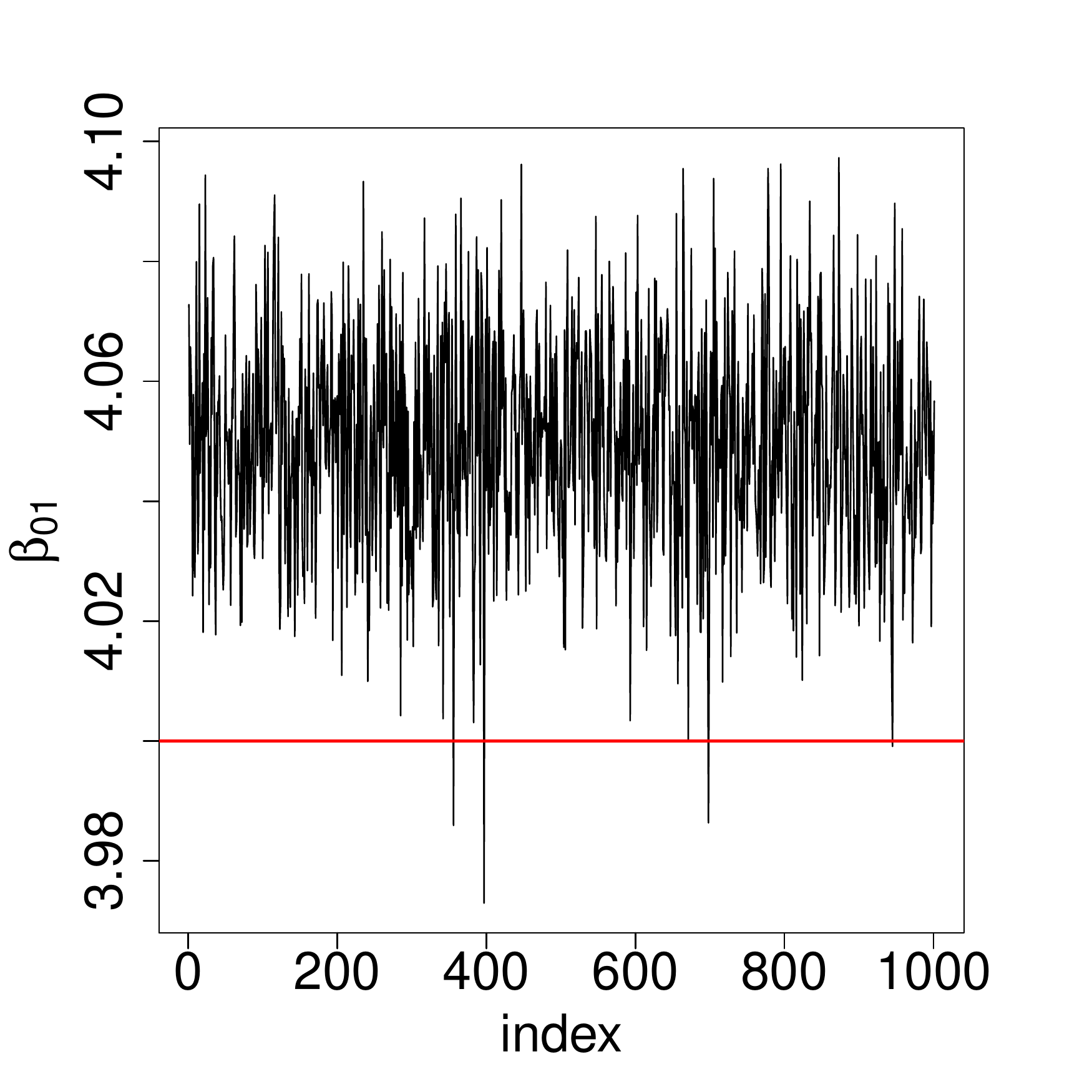} &
        \includegraphics[width=4cm]{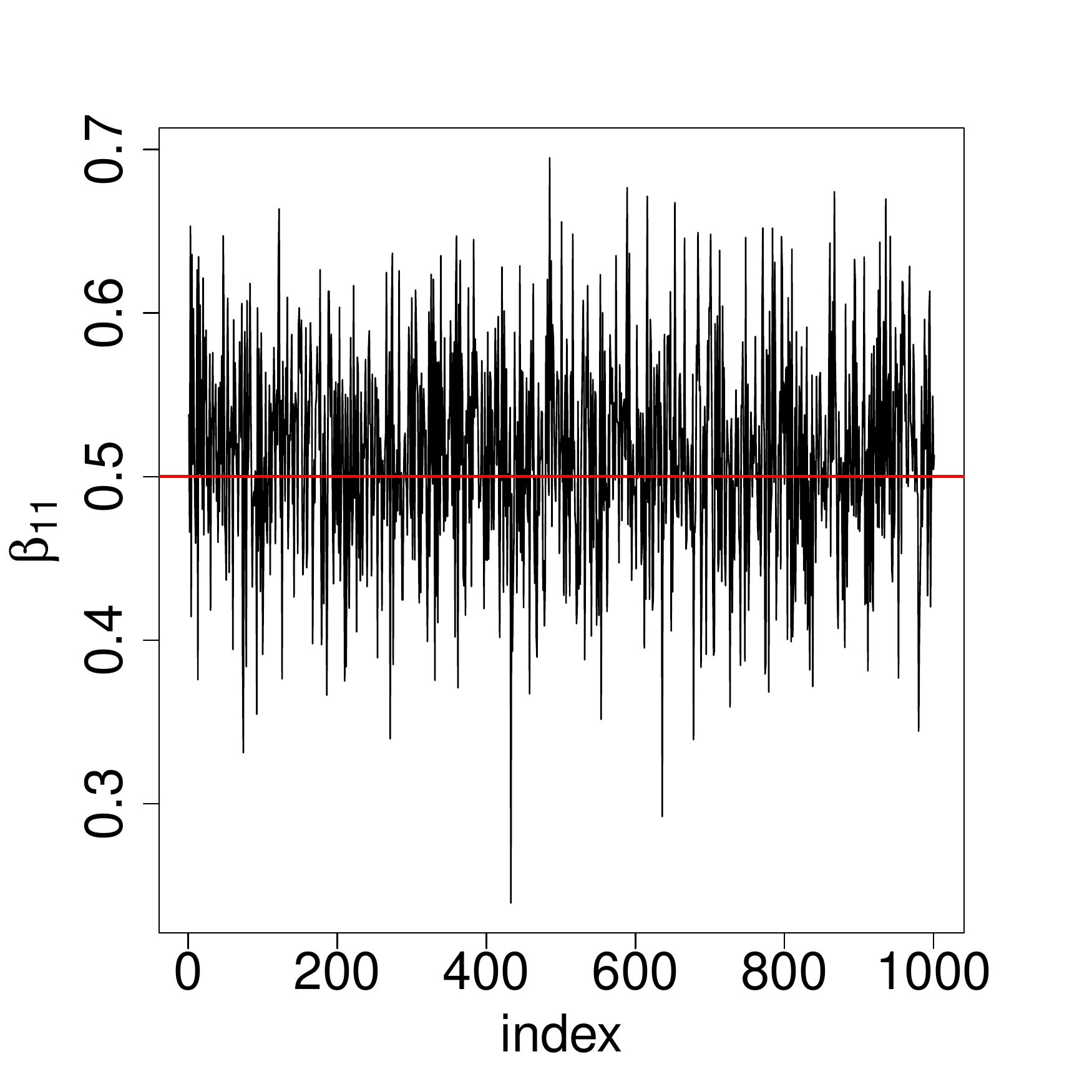} &
         \includegraphics[width=4cm]{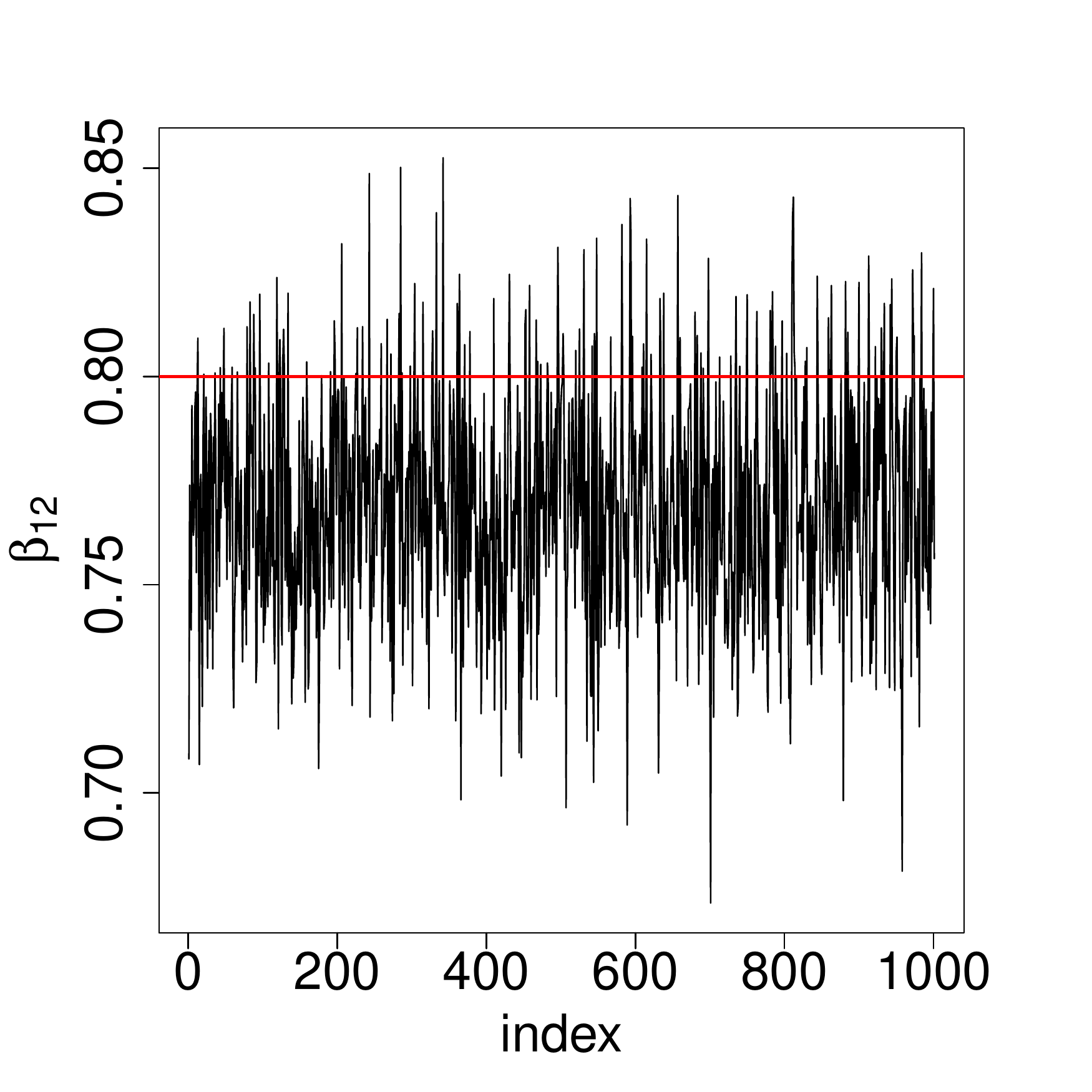} \\
       (a)  $\beta_{0,j=1}$ &  (b) $\beta_{0,j=2}$ &   (c) $\beta_{1,j=1}$ & (d)  $\beta_{1,j=2}$\\
         \includegraphics[width=4cm]{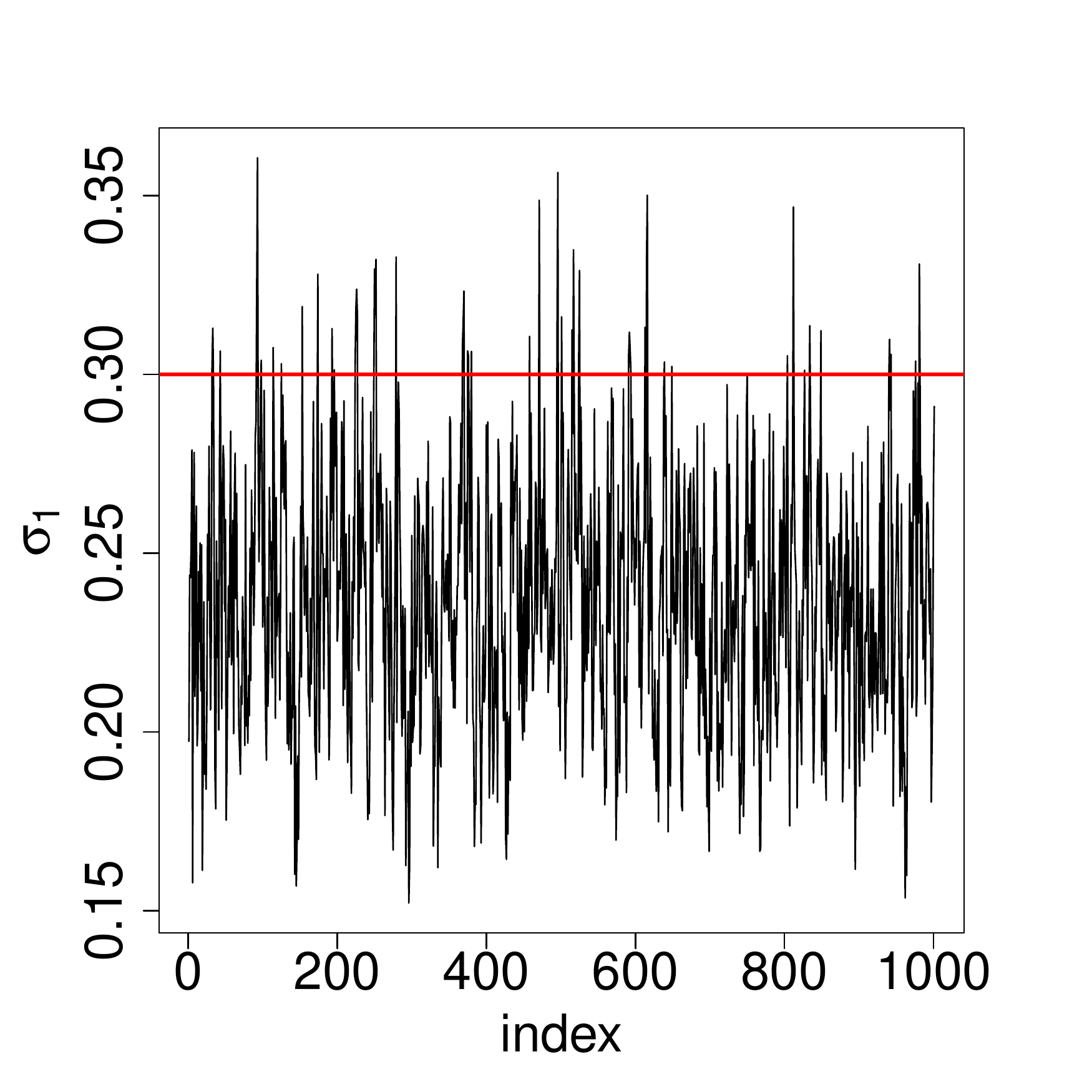} &
       \includegraphics[width=4cm]{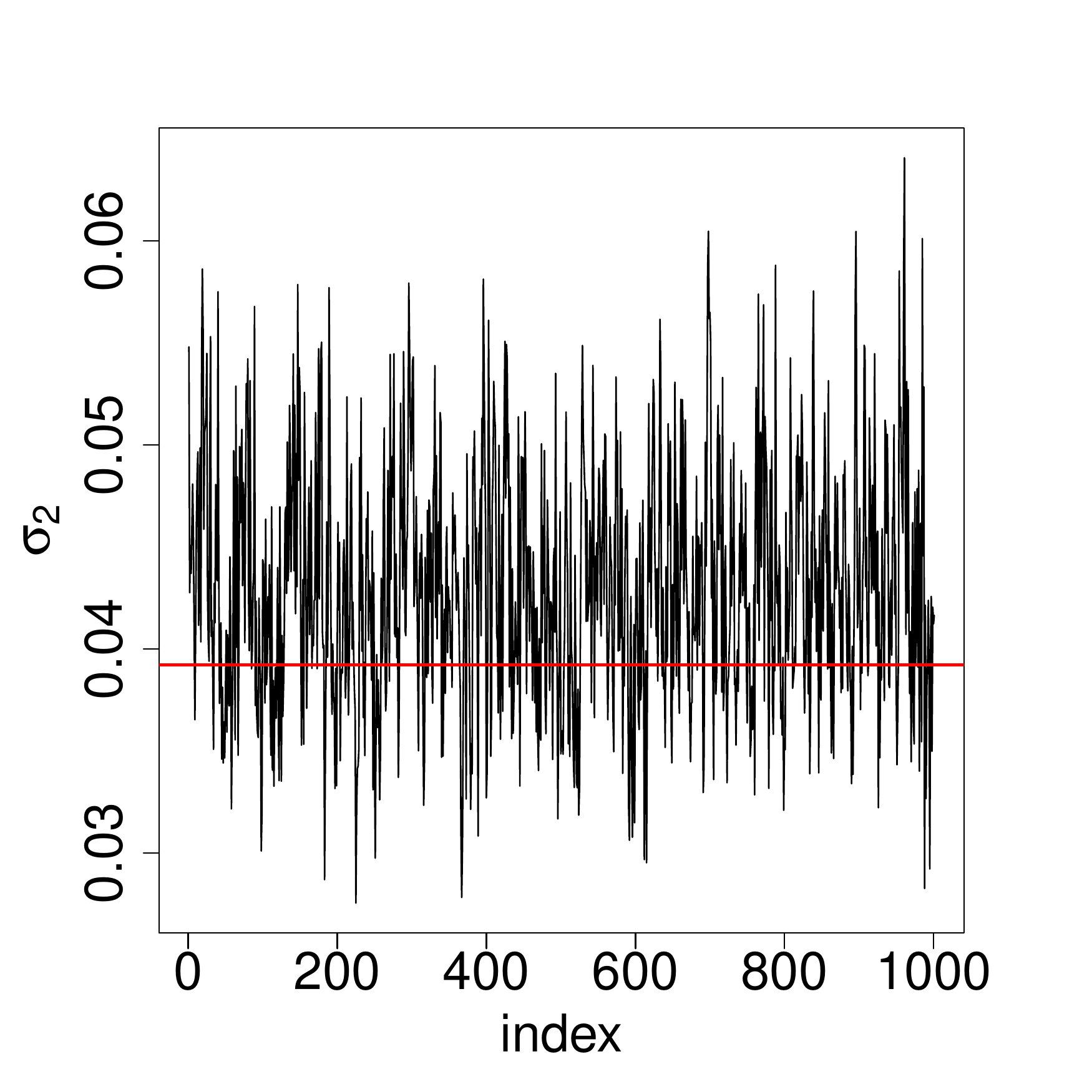} &
        \includegraphics[width=4cm]{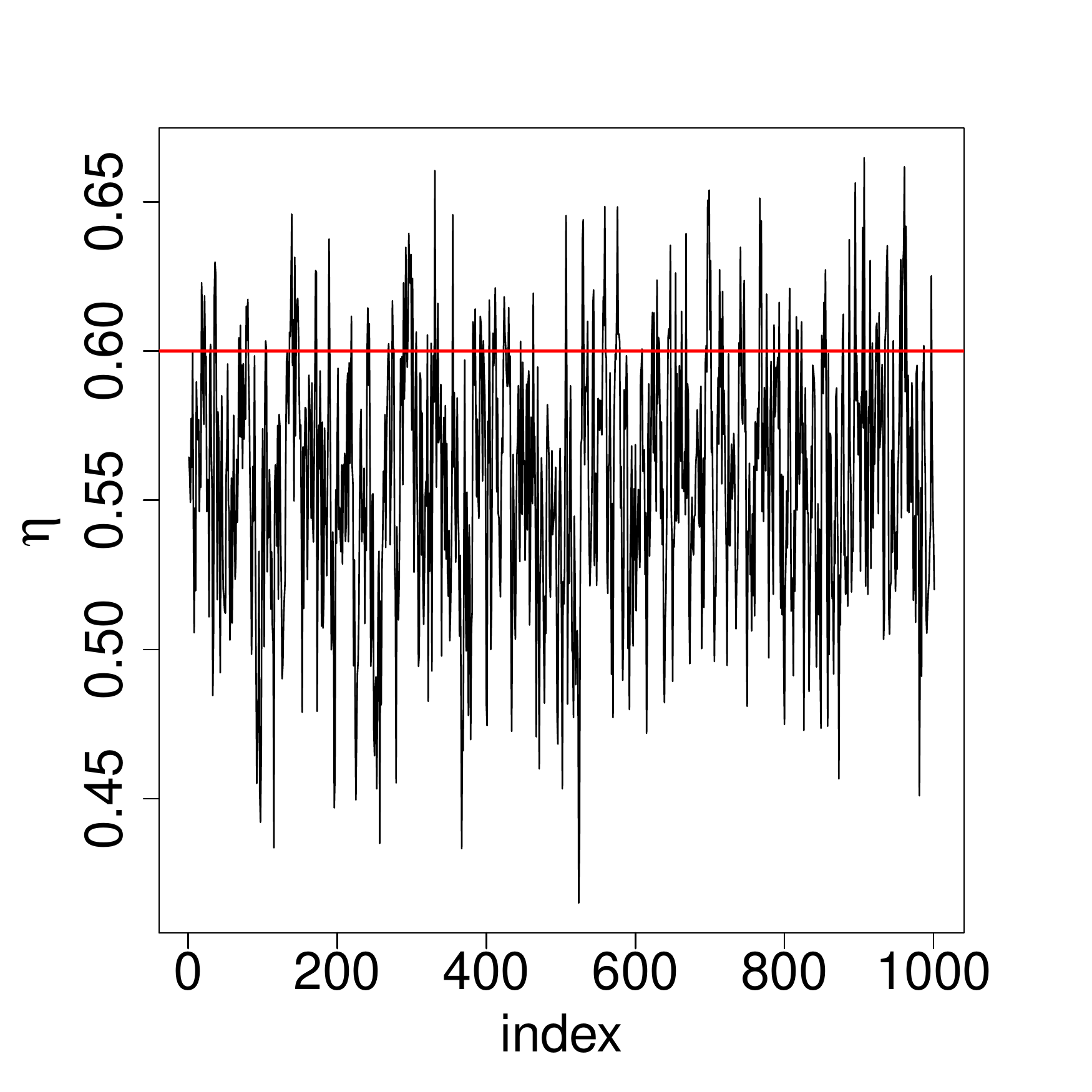} &
         \includegraphics[width=4cm]{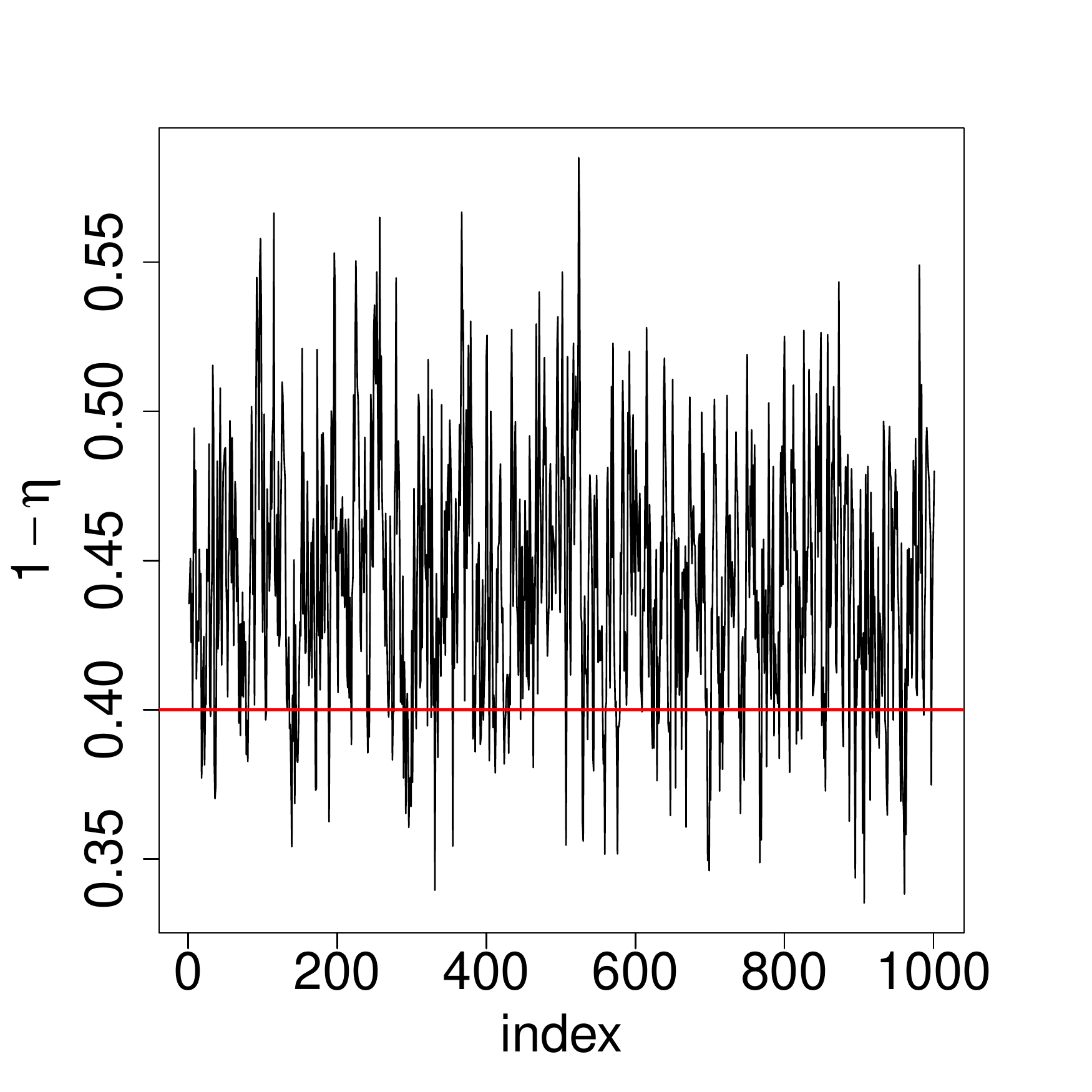}\\ \\
         (e) $\sigma^2_{1}$ & (f) $\sigma^2_{2}$ & (g) $\eta$ & (h) $1-\eta$ \\
    \end{tabular}
\end{center}
\caption{Simulated data set: trace plot of posterior parameters $\beta$, $\sigma$ and $\eta$ via data augmentation for the mixture survival model. Red lines presents the true values. } \label{fig2apB}
\end{figure}

\begin{figure}[H]
\begin{center}
    \begin{tabular}{cccc}
      \includegraphics[width=4cm]{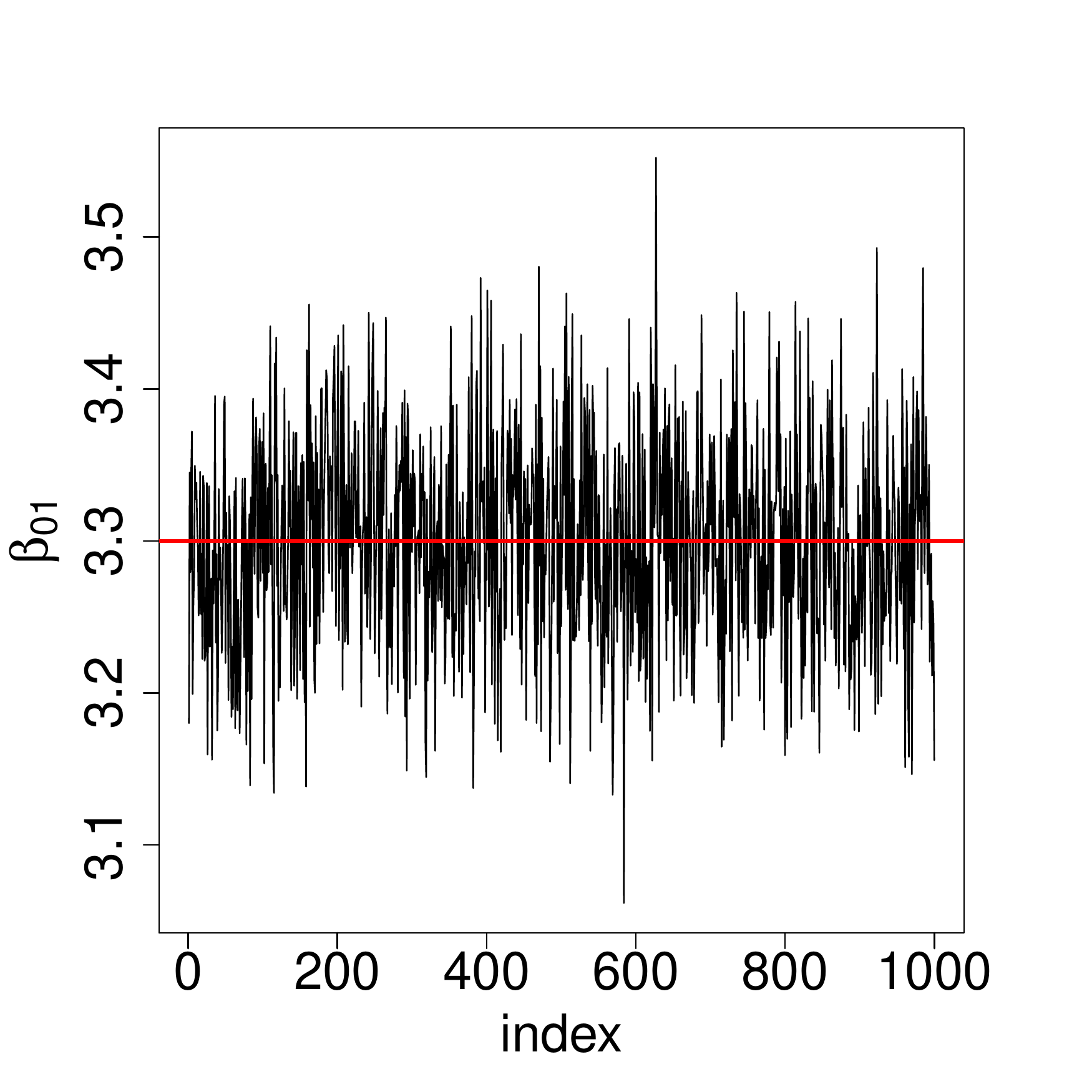} &
       \includegraphics[width=4cm]{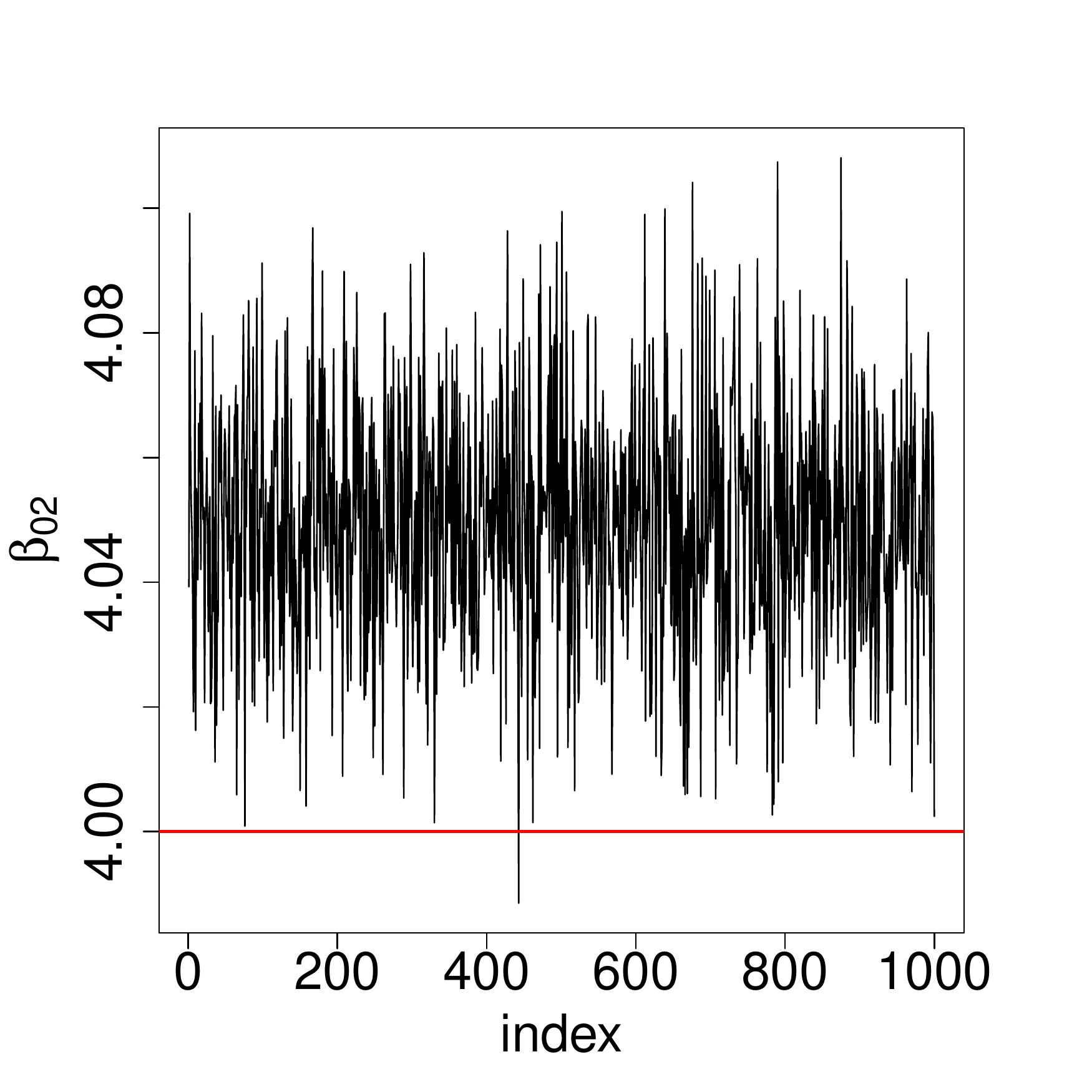} &
        \includegraphics[width=4cm]{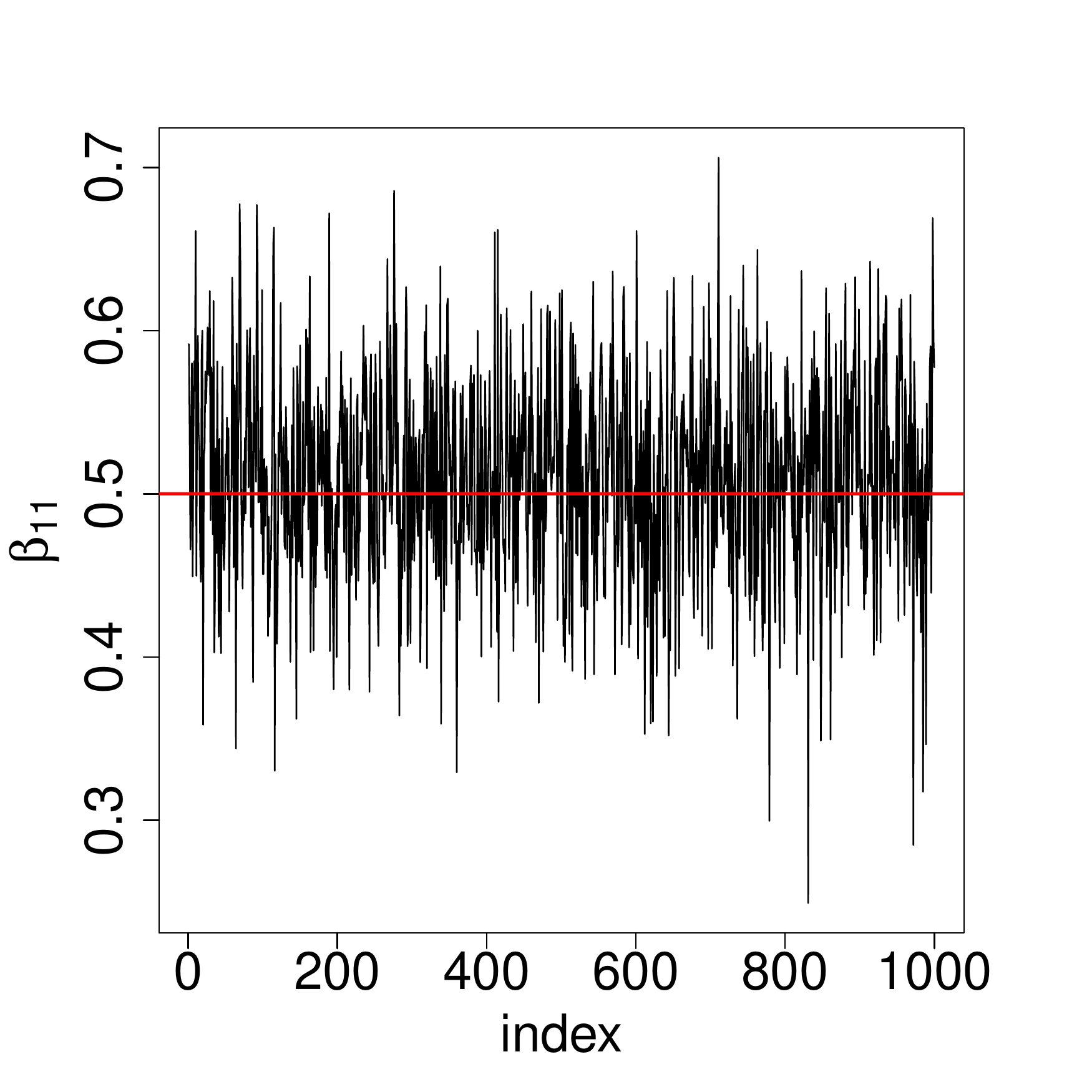} &
         \includegraphics[width=4cm]{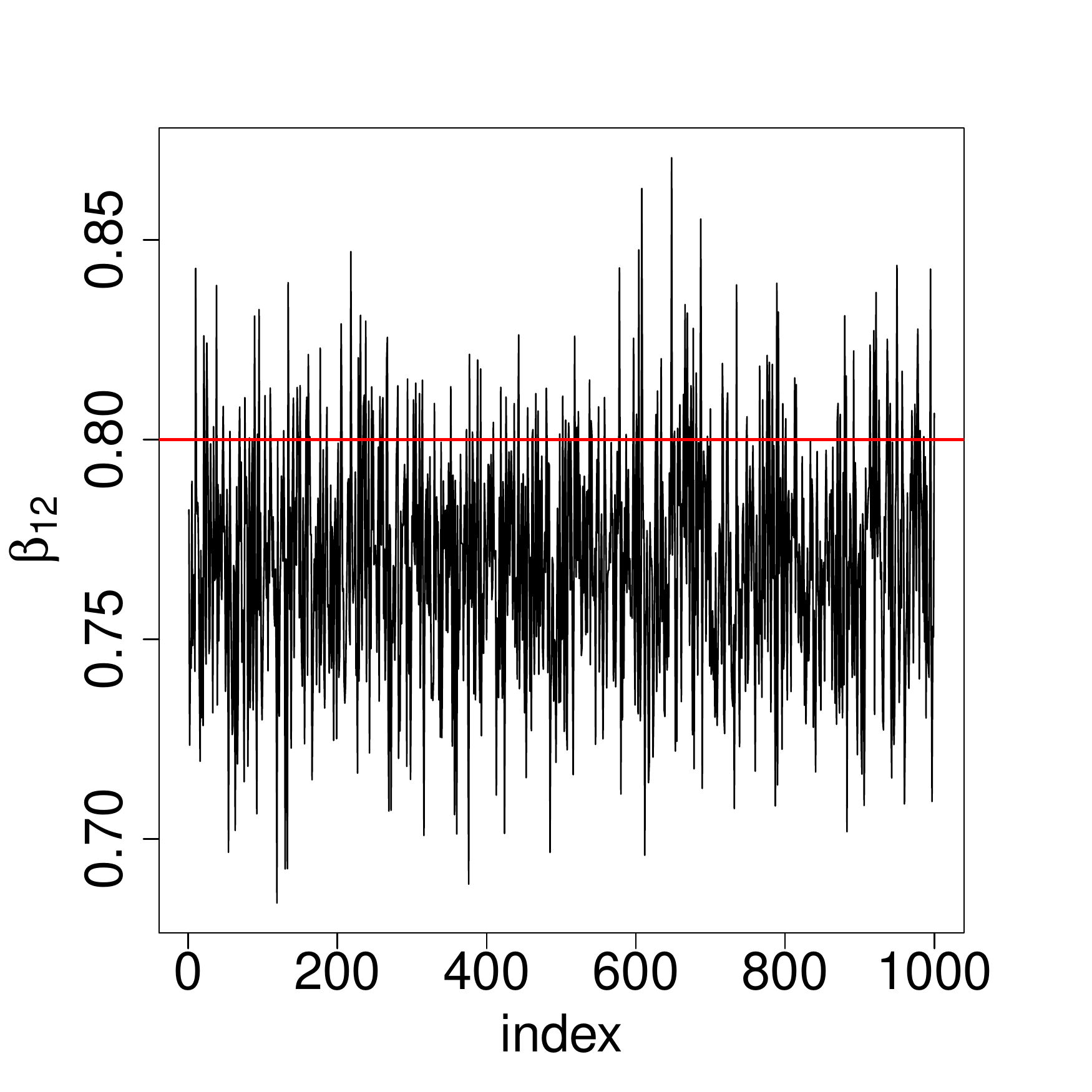} \\
       (a)  $\beta_{0,j=1}$ &  (b) $\beta_{0,j=2}$ &   (c) $\beta_{1,j=1}$ & (d)  $\beta_{1,j=2}$\\
         \includegraphics[width=4cm]{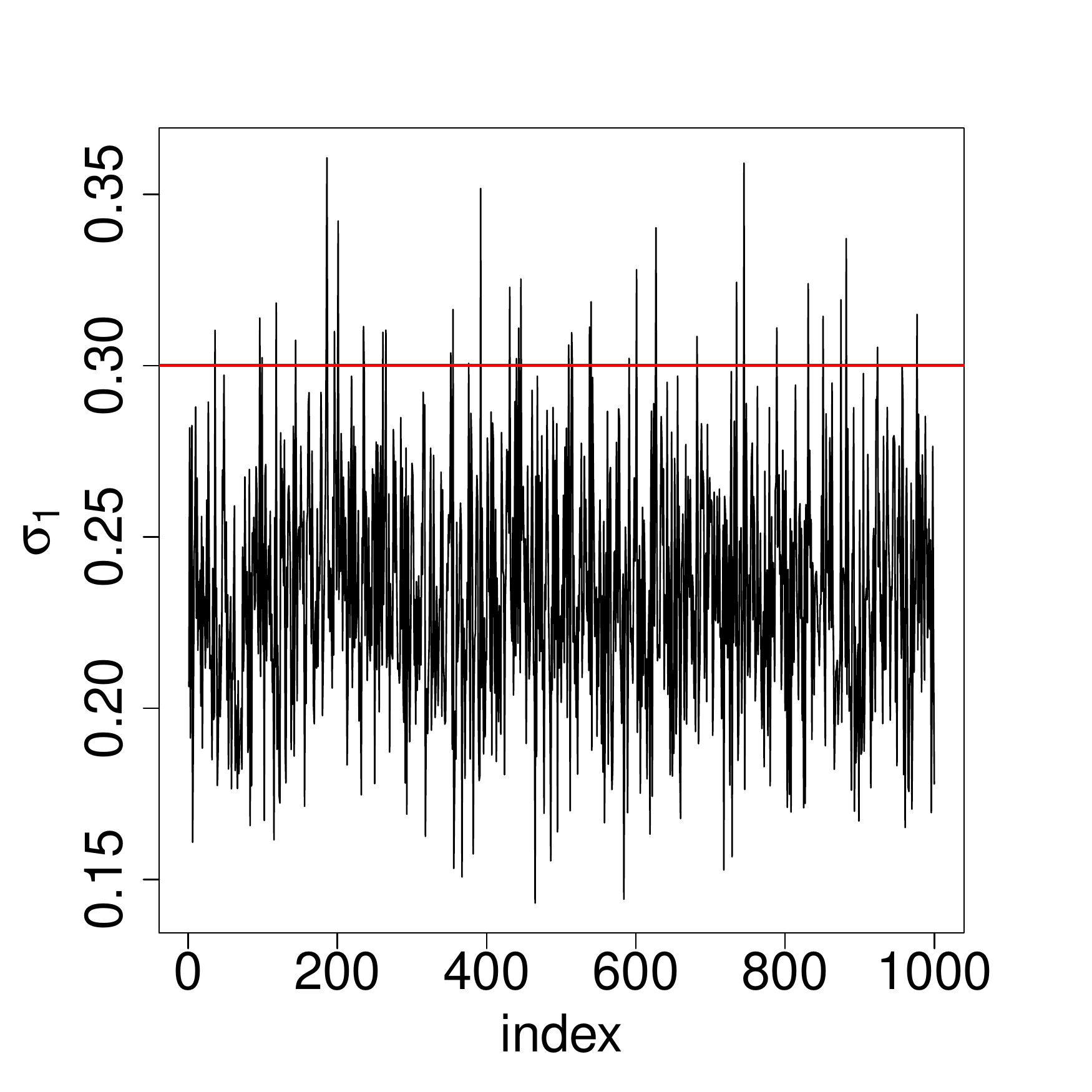} &
       \includegraphics[width=4cm]{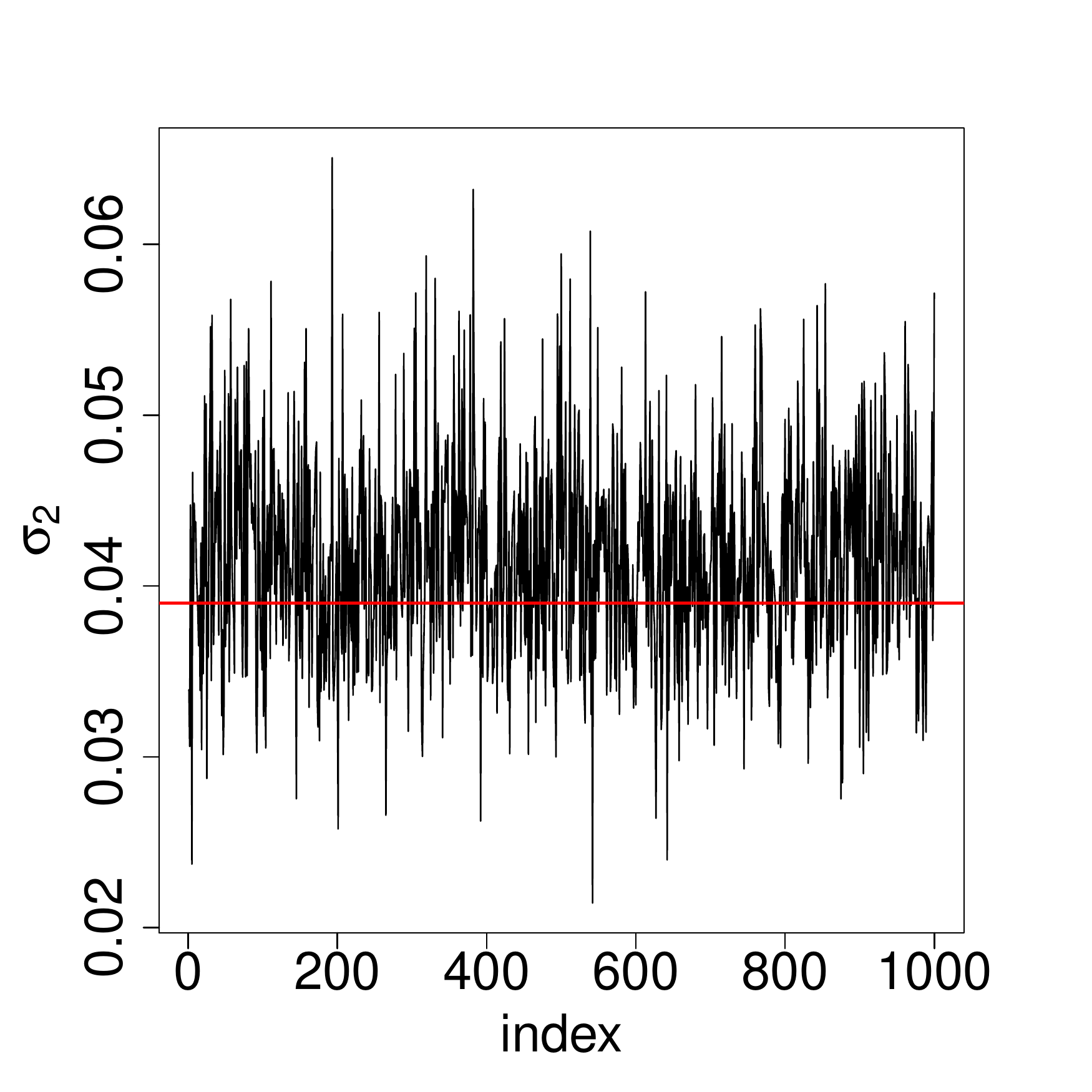} &
        \includegraphics[width=4cm]{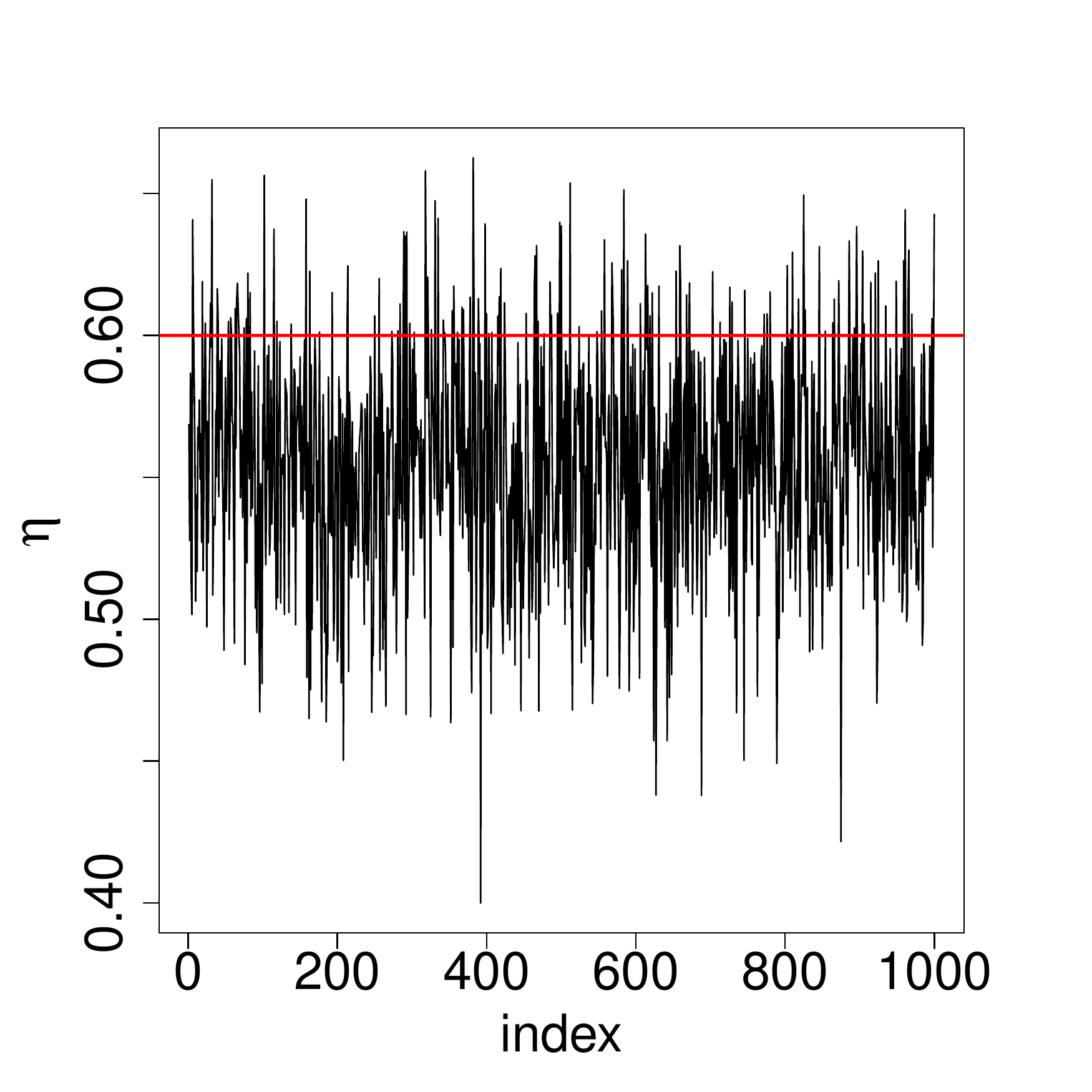} &
         \includegraphics[width=4cm]{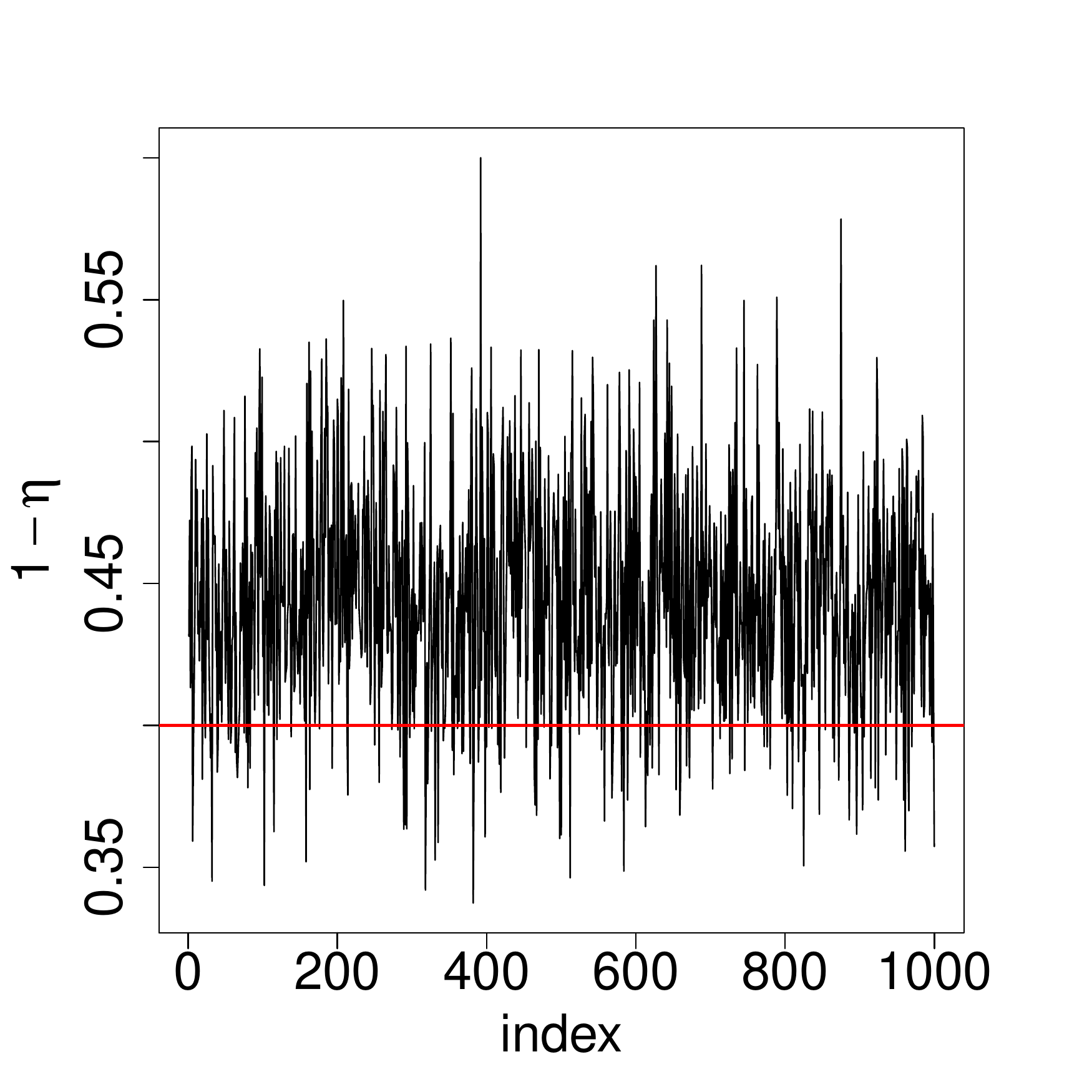}\\ \\
         (e) $\sigma^2_{1}$ & (f) $\sigma^2_{2}$ & (g) $\eta$ & (h) $1-\eta$ \\
    \end{tabular}
\end{center}
\caption{Simulated data set: trace plot of parameters $\beta$, $\sigma$ and $\eta$ via Stan for the mixture survival model. Red line presents the true values.} \label{fig3apB}
\end{figure}

Figures \ref{fig2apB} and \ref{fig3apB} present trace plots of the posterior parameters from the estimation process based on artificial data presented scenario (40\% censored data, $n=1,000$ and $\eta=0.6$) for both the Gibbs sampler with data augmentation and Stan. The chains present a stationary behaviour and contemplate the true values of parameters. The good performance of the survival posterior curves is shown in Section \ref{sec3.1}.

\bibliographystyle{apalike}
\bibliography{references}
\end{document}